\documentclass[a4paper, 11pt]{article}
\usepackage{my_jcap}

\usepackage{subcaption}  

\usepackage{natbib}
\setcitestyle{aysep={}}
\usepackage{graphicx}	
\usepackage{amsmath}	
\usepackage{breqn}

\usepackage{fontawesome5}
\usepackage{color}
\usepackage{xcolor}

\allowdisplaybreaks  

\usepackage{mdframed}
\newmdenv[
    linecolor=gray,            
    backgroundcolor=gray!10,    
    linewidth=0pt,            
    roundcorner=0pt,            
    skipabove=\topsep,          
    skipbelow=\topsep,          
    innertopmargin=-5pt,         
    innerbottommargin=10pt,
    innerleftmargin=5pt,
    innerrightmargin=5pt,
    splitbottomskip=0pt,        
    splittopskip=30pt,           
    nobreak=false,               
]{mybox}

\newcommand{\kms}{{\rm \, km\hskip 1pt s}\ensuremath{^{-1}}}

\newcommand{\msun}{\ensuremath{\, {\rm M}_\odot}} 
         
\newcommand{\mpc}{\ensuremath{\, {\rm Mpc}}}         
\newcommand{\gpc}{\ensuremath{\, {\rm Gpc}}}
     
\newcommand{\hinvmpc}{\ensuremath{\, h{\rm \,Mpc^{-1}}}}

\newcommand{\eg}{{\sl e.g.},\hskip 1pt}

\newcommand{\nhat}{\hat n}

\newcommand*{\vcenteredhbox}[1]{\begingroup
\setbox0=\hbox{#1}\parbox{\wd0}{\box0}\endgroup}

\newcommand{\OrcidID}[1]{ \href[urlcolor = red]{https://orcid.org/#1}{\textcolor{lightgray}{\faOrcid}}}
\newcommand{\OrcidIDName}[2]{\href{https://orcid.org/#1}{#2}}

\newcommand{\Rtwohc}{R_{\rm 200c}}
\newcommand{\Mtwohc}{M_{\rm 200c}}

\newcommand{\fNL}{f_{\rm NL}}

\newcommand{\kOne}{k_1}
\newcommand{\kTwo}{k_2}
\newcommand{\kThree}{k_3}

\newcommand{\kvec}{\vec{k}}

\newcommand{\FFT}{\text{FFT}\,}
\newcommand{\iFFT}{\text{iFFT}\,}

\defcitealias{Scoccimarro2012PNGs}{S12}
\defcitealias{Sohn:2023:CMBbest}{S23}
\defcitealias{Sohn:2024:Colliders}{S24}
\defcitealias{paper1}{\textsc{Paper I}}
\defcitealias{paper2}{\textsc{Paper II}}
\defcitealias{paper3}{\textsc{Paper III}}

\usepackage[T1]{fontenc}   
\usepackage{lmodern}       
\usepackage{anyfontsize}   

\title{\fontsize{19.5pt}{24pt}\selectfont Primordial Physics in the Nonlinear Universe: \\ 
mapping cosmological collider models to\\
weak-lensing observables}

\author[1, 2, 3]{\OrcidIDName{0000-0003-3312-909X}{Dhayaa Anbajagane}
(\vcenteredhbox{\includegraphics[height=1.2\fontcharht\font`\B]{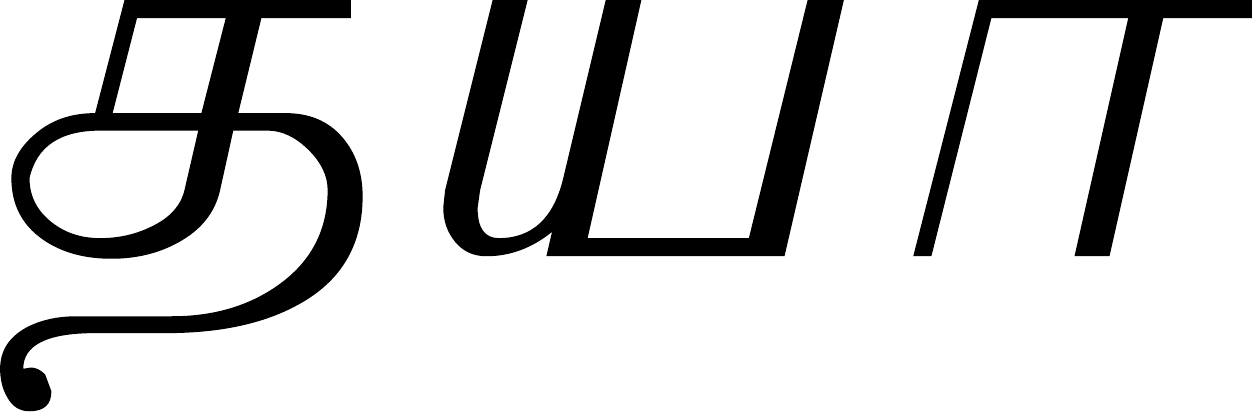}})}
\author[2, 4]{and \OrcidIDName{0000-0002-7577-0806}{Hayden Lee}}

\affiliation[1]{Department of Astronomy and Astrophysics, University of Chicago, Chicago, IL 60637, USA}
\affiliation[2]{Kavli Institute for Cosmological Physics, University of Chicago, Chicago, IL 60637, USA}
\affiliation[3]{NSF-Simons AI Institute for the Sky (SkAI), 172 E. Chestnut St., Chicago, IL 60611, USA}
\affiliation[4]{Center for Particle Cosmology, Department of Physics and Astronomy, University of Pennsylvania, Philadelphia, PA 19104, USA}
\emailAdd{dhayaa@uchicago.edu}
\emailAdd{haydenhl@sas.upenn.edu}

\abstract{Primordial non-Gaussianities (PNGs) are features in the initial density field that provide a window into the nonlinear dynamics of particles during the inflationary epoch. Among them, a distinctive set of signatures from ``cosmological collider physics'' originates through interactions of the inflaton with heavy particles active at high energies. The amplitude and form of these signatures depend on the strength and nature of the interactions. The corresponding features in large-scale structure have been studied predominantly through the use of perturbation theory, restricted to the linear regime of the density field. In this work, we implement a method for running cosmological simulations with arbitrary bispectra signals in their initial density field, and produce a simulation suite of over thirty PNG-generating templates, resolving the corresponding collider signatures in the strongly nonlinear regime of the density field. We detail the signals in a variety of late-time measurements---the matter power spectra, matter bispectra, the halo abundance, and halo bias. We then forecast the potential constraints on the signal amplitudes using weak lensing measurements from the Year-10 dataset of the Vera C. Rubin Observatory's Legacy Survey of Space and Time (LSST). The second and third moments of the lensing convergence field produce constraints that are competitive and complementary to those from the Cosmic Microwave Background. The data products are publicly released as part of the \textsc{Ulagam} simulation suite.
Our initial conditions generator is also publicly available at \url{https://github.com/DhayaaAnbajagane/Aarambam}.
}

\makeatletter
\def\@fpheader{\ }
\makeatother

\begin{document}

\maketitle
\flushbottom



\section{Introduction}

The physics of the earliest epochs in our Universe is rich in its phenomenology and is foundational to our understanding of the Universe's origin and evolution~\citep{Chen2010PNGReview, Achucarro2022InflationReview}. Yet, many aspects of this epoch remain elusive. 
Over the past decades, observations of the Cosmic Microwave Background (CMB) have provided a remarkably precise snapshot of the early Universe~\citep{Planck:2014:PNGs, Planck:2016:PNGs, Planck2020PNGs}, enabled by multiple high-accuracy experiments~\citep{Carlstrom2011, ACT:2016, Planck:2020:LegacyOverview}. These results have established inflation as the leading paradigm for the early Universe, in which the cosmos undergoes a brief period of rapid accelerated expansion~\citep{Guth:1981:Inflation, Linde:1982:Inflation, Guth2004Inflation}.

In its simplest form, inflation is driven by a single scalar field---the inflaton---slowly rolling down a nearly flat potential and exhibiting weak self-interactions. 
In this case, the statistics of the primordial density field are nearly Gaussian. 
However, the presence of additional fields or nontrivial inflaton interactions can lead to departures from this Gaussianity, known as primordial non-Gaussianity (PNG). The leading signatures of PNG are captured by the bispectrum, i.e., correlations among three Fourier modes of the density field, with an amplitude conventionally parameterized by $\fNL$ \citep[see][for a review]{Chen2010PNGReview}.

Three canonical types of primordial non-Gaussianity---local, equilateral, and orthogonal---are often used as benchmarks in analyses of cosmological surveys; see~\citet{Senatore2010WMAP5pngs} for details on the corresponding templates. 
However, a much broader landscape of possible signatures can arise from self-interactions of the inflaton or its couplings to additional particles~\citep{Chen:2010:QSF,Baumann:2011nk,Assassi:2012zq}. 
Since inflation may have occurred at energy scales as high as $10^{15}\,{\rm GeV}$, such interactions could involve particles far too massive to be produced in terrestrial experiments. 
This makes the inflationary epoch a unique laboratory for probing fundamental physics at otherwise inaccessible energies.
The resulting research program, often referred to as ``cosmological collider physics''~\citep{Nima:2015:Colliders}, has seen significant progress in recent years, particularly on the theoretical modeling side \citep{Lee:2016vti,Flauger:2016idt,Chen:2016hrz,Kumar:2017ecc,Bordin:2018pca,Arkani-Hamed:2018kmz,Baumann:2019oyu,Wang:2019gbi}, but now with growing advancements on the observational front as well \citep{Sohn:2024:Colliders, Cabass:2025:Colliders, Philcox:2025:PaperIII}.

The landscape of inflation constraints is currently dominated by observations of the CMB. 
Measurements of three-point and four-point correlations in the temperature and polarization fields have placed the strongest available constraints on many inflationary models, including some cosmological collider models \citep{Planck:2014:PNGs, Planck:2016:PNGs, Planck2020PNGs, Sohn:2024:Colliders, Philcox:2025:PaperIII}.
However, existing CMB measurements are already saturating the available information for constraining $\fNL$~\citep{Achucarro2022InflationReview}. 
The uncertainty on measurements of correlation functions (or polyspectra in Fourier/harmonic space) scales with the number of available modes in a given dataset: 
for two-dimensional fields, this number scales with area, while for three-dimensional fields, it scales with volume. 
The CMB measurements from \textit{Planck} are already mature in their control of systematics and access the widest area of the sky possible (while accounting for loss in area due to galactic dust, point sources, etc.). 
A next-generation, cosmic-variance limited experiment will improve on existing $\fNL$ constraints by a factor of 2 (5) for the equilateral and orthogonal (local) types \citep[][see their Figure 5]{Achucarro2022InflationReview}.

In light of this, the community has also focused on using the late-time Universe as a probe of inflationary physics~\citep{Achucarro2022InflationReview}. 
Crucially, late-Universe observables trace a three-dimensional field and therefore access a much larger number of modes, offering the potential for significantly improved constraints on $\fNL$. 
This has motivated extensive efforts to exploit the spatial clustering of galaxies---through measurements of the two- and three-point correlation functions---to constrain PNG~\citep{Cabass2022MultifieldBOSS, Cabass2022SingleFieldBOSS, Damico2022BossPNG, Philcox2022BossPNG, Cabass:2025:Colliders}. 
In addition, weak lensing measurements---which are often obtained in the same surveys that map galaxy positions--have been shown to provide constraints that are both competitive and complementary to those from galaxy clustering~\citep{Anbajagane2023Inflation}. 
Further promising probes include the kinematic Sunyaev-Zel'dovich effect, which traces large-scale velocity fields in the Universe~\citep[for a review]{Carlstrom2002SZReview}, and is expected to deliver precise PNG constraints in the near future~\citep[\eg][]{Munchmeyer:2019:kSZ_fNL}.

It is important to note that all these efforts above---and particularly the existing constraints from observations---rely on perturbative approaches to modeling inflationary signatures in the density field. 
While these approaches are robust and flexible, they are necessarily restricted to the quasi-linear regime, where the matter over-density satisfies $\delta \rho_m / \bar{\rho}_m \ll 1$. 
Consequently, such models are unable to access the nonlinear regime of the density field, which is known to contain significant cosmological information \citep[see][for reviews on some nonlinear probes]{Schneider:2005:Lensing, Allen:2011:Clusters}. 
As a result, this latter regime has not been utilized in constraints on $\fNL$. 
However, as discussed in \citet{Anbajagane2023Inflation}, the landscape of inflation analyses has now changed both due to the development of algorithms that can generate initial conditions with PNGs for use in simulations of large-scale structure and the development of computing infrastructure that now enables simulation suites to produce $\mathcal{O}(10^3 - 10^4)$ simulations. 
Thus, it is paramount to explore the sensitivity of nonlinear regime of the structure formation to the particle physics interactions of the early Universe.

Throughout this work, we consider inflationary signatures encoded in the bispectrum. 
Naively, generating a three-dimensional density field with a given bispectrum requires a three-dimensional convolutional integral per volume element of the density field, which for a standard simulation of $N_{\rm grid} = 2048$ necessitates $N_{\rm grid}^3 \times N_{\rm grid}^3 \sim \mathcal{O}(10^{18} - 10^{20})$ floating-point calculations. 
To overcome this computational barrier, \citet[][henceforth \citetalias{Scoccimarro2012PNGs}]{Scoccimarro2012PNGs} introduced an algorithm that exploits factorization of a given bispectrum to then reduce the three-dimensional integral into three one-dimensional integrals, resulting in orders-of-magnitude improvements in the computational complexity of the algorithm. Many works have since utilized this approach for generating initial conditions \citep{Coulton2022QuijotePNG, Jung2023fNLHMFQuijote, Goldstein:2024:CosmoColl, Goldstein:2025:CosmoColl}. 

Existing simulations have mainly focused on the three types of PNGs: local, equilateral, or orthogonal \citep[\eg][]{ Coulton2022QuijotePNG, Anbajagane2023Inflation, Boryana:2024:AbacusPNG, Adame:2024:PNGUnit}.
By contrast, cosmological collider models, which have a rich variety in the bispectra generated by particle interactions, have so far not been simulated.
This is because their bispectra are non-separable and are therefore unable to be directly simulated using the algorithm from \citetalias{Scoccimarro2012PNGs}.\footnote{One notable exception is the work of \citet{Goldstein:2024:CosmoColl, Goldstein:2025:CosmoColl}, who construct by-hand a separable kernel of a collider-inspired model that is valid for a specific (squeezed) limit of the bispectrum. Our work generalizes this to the full domain of the bispectrum and focuses on the original collider templates. We discuss the technical details of these differences further in Section \ref{sec:sims:ICs}.} However, \citet[][henceforth \citetalias{Sohn:2023:CMBbest}]{Sohn:2023:CMBbest} and \citet[][henceforth \citetalias{Sohn:2024:Colliders}]{Sohn:2024:Colliders} introduced an approach for decomposing any bispectrum into a set of basis functions, such that a non-separable bispectrum can be approximated by a sum of many separable pieces. 
Indeed, such an approach was already discussed in Appendix D of \citetalias{Scoccimarro2012PNGs}.
The challenge then lies in choosing a basis that is favorable for both theoretical accuracy and numerical stability. 
In this work, we build on the basis decomposition method of \citetalias{Sohn:2023:CMBbest}, and extend it by explicitly constructing basis functions that are tailored to the algorithm of \citetalias{Scoccimarro2012PNGs}. 
In particular, we explicitly choose functions that prevent infrared divergences in the power spectrum (see Section \ref{sec:sims:ICs}). 
With this approach, we are now able to simulate any bispectrum template, as long as the template can be accurately decomposed into the chosen basis functions.\footnote{As we will discuss later, the choice of basis functions is the most critical decision in this work and is also the aspect of our method with the largest potential for improvement.}

In this work, we present the first simulation suite to include the full cosmological collider bispectra in the initial conditions. 
We simulate over thirty templates and study the signatures of these bispectra in the strongly nonlinear regime of the density field. In particular, we analyze their impact on the late-time matter power spectrum and bispectrum, the halo mass function and halo bias, and then forecast the sensitivity of weak-lensing measurements to these signatures.

This paper is organized as follows: Section~\ref{sec:sims} discusses the method for adding bispectrum signatures to the initial conditions, and the bispectrum models we consider in this work. Section~\ref{sec:results} presents the signatures of PNGs in the matter field and halo field, and forecasts the constraints obtained from upcoming weak-lensing observations. We summarize in Section~\ref{sec:conclusions}. Appendix \ref{appx:ICs} provides other technical details in simulating the initial conditions, while Appendix \ref{appx:Validation} presents a full validation of our method for all PNG models considered in this work.

\section{Simulations}\label{sec:sims}

Our simulations are part of the \textsc{Ulagam} suite \citep{Anbajagane2023Inflation}, which contain full-sky lightcones built using the \textsc{PkdGrav3} $N$-body code \citep{Potter2017Pkdgrav3}. The $N$-body simulations of this work follow the same procedure from \citet{Anbajagane2023Inflation}, with one key alteration in how the initial conditions are generated. The validation of the simulation methodology can also be found in that work. We will describe further below our procedure for generating said initial conditions (Section \ref{sec:sims:ICs}), and then introduce the inflation models considered in this work (Section \ref{sec:sims:Models}).

The $N$-body simulations are run with $512^3$ particles, within volumes of $V = L^3 = (1 \gpc/h)^3$ with $h = H_0/(100 \kms/\mpc)$ being the dimensionless Hubble constant. The displacements used in the initial conditions are generated using 2nd-order Lagrangian Perturbation Theory (2LPT). We follow \citet{Navarro2020Quijote, Coulton2022QuijotePNG} in generating the matter transfer function at $z = 0$ using \textsc{CAMB} and rescaling it to our starting redshift, $z = 127$, using the analytic growth rate in matter-dominated Universes. We produce 100 snapshots between $z = 127$ and $z = 0$, and \textsc{PkdGrav3} also internally produces lightcone shells for the redshift of each snapshot, duplicating the 3D volumes as needed to extend the observable volume to the one encompassed within a given redshift. Such \textsc{PkdGrav3}-generated lightcones have been used extensively in simulation-based analyses of weak-lensing \citep[\eg][]{Fluri2019DeepLearningKIDS, Fluri2022wCDMKIDS, Gatti:2024:WPH, Jeffrey:2025:Likelihood, Prat:2025:Homology}. Halos are identified in each snapshot using both a position-based Friends-of-Friends algorithm internal to \textsc{PkdGrav3} as well as by explicitly running \textsc{Rockstar} \citep{Behroozi2013Rockstar} on the snapshots. The latter is a new addition to our pipeline relative to that used in \citet{Anbajagane2023Inflation} and enables more robust quantification of the statistics of halos in our simulations. We therefore use the latter catalogs in all results presented in this work. We use the halo mass/radius definition $\Mtwohc = 200 \rho_c \times 4\pi \Rtwohc^3$, where $\rho_c$ is the critical density at a given epoch. 

As previously mentioned, Appendix A of \citet{Anbajagane2023Inflation} provides a validation of the simulation pipeline. In particular, Figure 10 in that work confirms the matter power spectrum is converged to 5\% (8\%) at $z = 0$ ($z = 0.6$), across the range $10^{-2} \leq k [h/{\rm Mpc}] \leq 1$. The lensing signal peaks at $z \sim 0.5$ and in our analysis below is sensitive to up to $k = 1 h/{\rm Mpc}$. Errors at the 10\% level do not impact the conclusions of our work. Figures 12 and 13 in \citet{Anbajagane2023Inflation} also validate the halo mass function and halo bias measured in our fiducial simulations, finding the quantities agrees with prior simulation models. Increasing the resolution of our simulation suite will boost the power on smaller scales (by $\approx 10\%$), thereby increasing our signal and improving constraints. Thus, the parameter forecasts presented in this work are a conservative estimate in this regard.

\subsection{Generating Initial Conditions}\label{sec:sims:ICs}

A bispectrum describes correlations between three wavevectors and can be written as
\begin{equation}\label{eqn:kernel}
    B(\kOne, \kTwo, \kThree) = 2\fNL K_{12}(\kOne, \kTwo, \kThree) P(\kOne) P(\kTwo) + 2 {\rm \,\, perm.}\,,
\end{equation}
where the form of the kernel $K_{12}$ depends on the inflationary model~\citep[\eg][]{Chen2010PNGReview, Sohn:2024:Colliders, Philcox:2025:PaperI}. The corresponding dimensionless shape function is defined by
\begin{equation}
    S(\kOne, \kTwo, \kThree) = (\kOne\kTwo\kThree)^2B(\kOne, \kTwo, \kThree)\,.
\end{equation}
We will sometimes write this as $S(\kOne, \kTwo, \kThree) = S(\kvec_1, \kvec_2)$, with $\kOne = |\kvec_1|$, $\kTwo = |\kvec_2|$, and $\kThree = |\kvec_1 + \kvec_2|$. Depending on the context, we will use $B$ and $S$ interchangeably in the discussions to follow.

To generate a non-Gaussian potential, $\phi_{\rm NG}$, we perform the following (convolution) integral of a given Gaussian potential, $\phi_G$:
\begin{equation}\label{eqn:conv}
    \phi_{\rm NG}(\kvec) = \phi_{\rm G}(\kvec) + \fNL\int d^3k_1 d^3k_2\, (2\pi)^3 \delta(\kvec - \kvec_{12}) K_{12}(\kvec_1, \kvec_2) \phi_{\rm G}(\kvec_1)\phi_{\rm G}(\kvec_2) 
    \,,
\end{equation}
where $\vec k_{12}  = \vec k_1 + \vec k_2$ and $ K_{12}(\kvec_1, \kvec_2)=K_{12}(\kOne, \kTwo, \kThree)$. 
This direct evaluation leads to a computational complexity of $N_{\rm grid}^6$.\footnote{The convolution integral scales as $N_{\rm grid}^3$ (the delta function reduces the naive $N_{\rm grid}^6$ complexity to $N_{\rm grid}^3$) and this integral must be done for each of the $N_{\rm grid}^3$ entries of the $\phi(\kvec)$ grid, leading to a scaling of $N_{\rm grid}^6$.} 
In \citetalias{Scoccimarro2012PNGs}, they showed that the integral in Eq.~\eqref{eqn:conv} can be done efficiently via three Fast Fourier Transforms (FFTs) \textit{as long as the kernel is separable} in $\kOne, \kTwo, \kThree$. That is, if we split the kernel as $K_{12}(\kOne, \kTwo, \kThree) = f_1(\kOne) f_2(\kTwo) f_3(\kThree)$ then the above integral can be evaluated as
\begin{equation}\label{eqn:sim:ICs:S12}
    \phi_{\rm NG}(\vec{k}) = \phi_{\rm G}(\vec{k}) + \fNL\bigg[f_3(k_3) \times \FFT\Big\{\iFFT \Big(f_1(k_1) \phi_G(k_1)\Big) \times \iFFT \Big(f_2(k_2)\phi_G(k_2)\Big)\!\Big\}\!\bigg],
\end{equation}
following the approach of \citetalias{Scoccimarro2012PNGs}. We denote the inverse FFT operation as $\iFFT = \FFT^{-1}$. The scaling of the algorithm in this case is $N_{\rm grid}^3 \log(N_{\rm grid})$. This is the methodology we employ in this work for generating non-Gaussian initial conditions. Following \citet{Navarro2020Quijote}, we generate $\phi_{\rm NG}(\vec{k})$ on a grid that has 8 times as elements as the particle counts, i.e. $N_{\rm grid} = 1024^3$ given we use $N_{\rm particle} = 512^3$ in this work, as this minimizes aliasing. \citet{Adame:2025:fNLAlias} also show the effect of aliasing is negligible when generating local-type PNGs.

To apply Eq.~\eqref{eqn:sim:ICs:S12}, we require a procedure for decomposing arbitrary kernels into such separable functions. 
Many methods have been introduced to decompose $S$ (which is equivalent to decomposing $K_{12}$) into a separable function of the kind, $S(\kOne, \kTwo, \kThree) = f(\kOne)g(\kTwo)h(\kThree)$, where $f,g,h$ are one-dimensional functions of wavenumbers \citep[][\citetalias{Sohn:2023:CMBbest}, \citetalias{Sohn:2024:Colliders}]{Komatsu:2005:KSW, Fergusson:2012:Modal}. 
These methods originated in the CMB literature, where they greatly simplify the projection integrals required to map 3D real-space bispectra onto 2D angular bispectra. 
Similar techniques have also been applied in the context of angular bispectra and trispectra in large-scale structure~\citep[\eg][]{Assassi:2017:CLs, Lee:2020:CLs, Chen:2021vba}. 
In this work, we adopt the method developed by \citetalias{Sohn:2023:CMBbest}, who applied it to a set of inflationary models that we also study in Section \ref{sec:sims:Models}. 

Following \citetalias{Sohn:2023:CMBbest}, we first expand the shape function as 
\begin{equation}\label{eqn:kernel:1}
    S(\kOne, \kTwo, \kThree) = \sum_{i,j,k = 0}^{N} \alpha_{ijk} \,q_i(\kOne) q_j(\kTwo) q_k(\kThree)\,,
\end{equation}
where $\alpha_{ijk}$ are coefficients and $q_i$ are basis functions. 
This approximates the target template as a sum of individually separable terms. 
The basis functions are defined as
\begin{align}\label{eqn:modefunc}
    q_i(k) = 
    \begin{cases}
      (k/k_\star)^{ (i - 3)x+2}
        & \text{if } i \leq 3\,,\\[3pt]
        \mathcal{P}_{i - 1}(\Tilde{k}) - A_{i - 1} & \text{if } i > 3\,,\\
    \end{cases}
\end{align}
where ${\cal P}_{i-1}$ is the Legendre polynomial of degree $i-1$ and the exponent $x = (4 - n_s)/3$ is included to account for the mild scale dependence of the primordial power spectrum.
Without loss of generality, we take $k_\star = 1 \,\,h{\rm /Mpc}$.\footnote{Other choices of normalization would simply rescale the derived coefficients, $\alpha_{ijk}$, and result in no changes to the approximation for a given template.} 
The quantity $\Tilde{k}$ is a normalized (dimensionless) wavenumber defined as 
\begin{equation}\label{eqn:kbar}
    \Tilde{k}  = -1 + 2\frac{k^x - k_{\rm min}^x}{k_{\rm max}^x - k_{\rm min}^x} \,.
\end{equation}
We set $k_{\rm min}$ and $k_{\rm max}$ based on the simulation configuration, and are given by $k_{\rm min} = \frac{2\pi}{L}$ and $k_{\rm max} = \frac{2\pi}{L} \frac{N_{\rm grid}}{2}$. Here, $L$ is the box size of the simulation and $N_{\rm grid}$ is the size of the grid used to generate the initial conditions. For the simulations in this work, we use $N_{\rm grid} = 1024$. The factor of two in the maximum wavenumber reflects the fact we only sample up to the Nyquist scale of the grid. 

The constant $A_{i - 1}$ in Eq.~\eqref{eqn:modefunc} is the coefficient associated with the $k^0$ term in the Legendre polynomial of degree $i - 1$. We explicitly subtract this constant factor from the basis function to prevent divergent behavior in the 1-loop power spectrum; we discuss this aspect further below and in Appendix \ref{appx:ICs:Divergent}. 
Note that Eq.~\eqref{eqn:modefunc} writes the polynomial in $\Tilde{k}$, not $k$. Therefore, $A_{i - 1}$ must be calculated by expanding the Legendre series into a polynomial in $k$ and identifying the coefficient of the $k^0$ term.\footnote{We evaluate this coefficient as $A_{i - 1} = P_{i - 1}(\Tilde{k}(k = 0))$, where $P$ are the Legendre polynomials and $\Tilde{k}(k = 0) = -1 - 2k_{\rm min}^x/(k_{\rm max}^x - k_{\rm min}^x)$ following Equation \eqref{eqn:kbar}.}

Our basis functions consist of four monomial terms that go from $k^{-1}$ to $k^2$, and Legendre polynomials of up to a user-defined order $N$. The four monomial terms allow the basis to exactly reproduce the shape functions of the local, equilateral, and orthogonal PNGs. 
As a result, our Legendre basis starts from the polynomial $\mathcal{P}_2$, thereby excluding the $\mathcal{P}_0(k) = 1$ and $\mathcal{P}_1(k) = k$ terms which are exactly degenerate with the monomial terms. 
The Legendre polynomials have attractive computational properties---primarily their numerical stability which arises from their orthogonal nature\footnote{The orthogonality properties of the polynomials are diminished due to the modification made in Eq.~\eqref{eqn:modefunc}. However, these modified polynomials are still more orthogonal than a simple monomial basis.}---and are therefore a reasonable choice of basis for this work. 
Compared to \citetalias{Sohn:2023:CMBbest}, we choose the argument of the Legendre functions to be $k^{(4 - n_s)/3}$ rather than just $k$; see Eq.~\eqref{eqn:modefunc}. This mimics the same choice made in defining the model templates (Section \ref{sec:sims:Models}). 
Our approach of using four monomial terms in the basis, rather than a single monomial term as in \citetalias{Sohn:2023:CMBbest}, is simply a choice and both approaches provide the same overall flexibility to the basis set. 
We use the latter as it allows the basis to \textit{exactly} match the local, equilateral, and orthogonal templates.

The decomposition of the shape function, $S$, into the basis, denoted as $\vec{Q}$, is done by solving a linear equation,
\begin{equation}\label{eqn:ICs:matrix_eqn}
    \sum_b\langle Q_a, Q_b \rangle \alpha_b = \langle Q_a, S \rangle\,,
\end{equation}
where $Q_a$ are the individual three-dimensional basis functions in the sum in Eq.~\eqref{eqn:kernel:1}, and we define the inner product as
\begin{equation}\label{eqn:ICs:innerproduct}
    \langle A, B \rangle = \int_T d^3\log k\,A(\vec{k})B(\vec{k})\,.
\end{equation}
We perform the (numerical) integral over $\log k$ instead of $k$ (where the latter was used in \citetalias{Sohn:2023:CMBbest}) to ensure that small-scales and large-scales contribute equally to the integral. 
Otherwise, the inner product is dominated by larger values of $k$ (small spatial scales) and the decomposition is less accurate for the low-$k$, large-scale regime of the template. 
The volume $T$ is the tetrapyd defined with $\kOne \geq \kTwo \geq \kThree$, $\kTwo + \kThree \geq \kOne$, and $k_{\rm min} \leq \kOne, \kTwo, \kThree \leq k_{\rm max}$~\citep{Sohn:2023:CMBbest}.

Following \citetalias{Sohn:2023:CMBbest}, the matrix in Eq.~\eqref{eqn:ICs:matrix_eqn} is solved iteratively using the conjugate gradient method in \textsc{Scipy} \citep{Virtanen2020Scipy}. We use up to $10^5$ iterations, or until the precision of the coefficients $\alpha_b$ achieves a fractional uncertainty of $10^{-12}$; in practice, we almost always achieve this uncertainty within $10^5$ iterations. We have checked the accuracy of our approximation both (i) quantitatively, using the inner product of the approximate template and the true template, and (ii) visually, by comparing the approximated template to the analytic one. For the former, we follow \citet{Planck:2014:PNGs}, which defines the quantity $\epsilon = \sqrt{2(1 - r^2)}$, with
\begin{equation}\label{eqn:ICs:eps}
    r = \frac{\langle \hat{S}, S \rangle}{\sqrt{\langle \hat{S}, \hat{S} \rangle \langle S, S \rangle}}\,,
\end{equation}
which quantifies the correlation between the target ($S$) and approximated ($\hat{S}$) template across the full Fourier-space volume. The quantity $\epsilon$ represents the predicted scatter between $\fNL$ constraints obtained from using the true analytic template and from using the approximated version \citep{Planck:2014:PNGs, Sohn:2023:CMBbest}. Following Appendix B of \citet{Planck:2014:PNGs} we see that for $r \sim 0.99$ we have $\epsilon = 0.2$, and so the $\fNL$ constraints using the approximated templates have a scatter of $\epsilon \times \sigma(\fNL) = 0.2 \times \sigma(\fNL)$ relative to the constraints from the true analytic template. Here, $\sigma(\fNL)$ is the uncertainty on the constraint when using the analytic template. In our case, most templates exhibit $r \gtrsim 0.999$, and all templates indicate $r \gtrsim 0.99$, indicating our approximations are adequate.

The necessity for high-precision, robust matrix inversion in solving Eq.~\eqref{eqn:ICs:matrix_eqn} is due to the highly correlated nature of our basis functions. Even though we have picked functions, $q_i$, that are mostly (but not exactly) orthogonal, this is true only in one dimension. Once we construct the three-dimensional basis functions, $Q_a$, each function is not orthogonal from the rest. Functions with higher mode indices are generally more highly correlated with each other. The technique of \citetalias{Sohn:2023:CMBbest} does not require an orthogonal basis, which makes it particularly powerful in its flexibility and enables redefining the basis functions to suit our specific needs (for example, to cancel divergent terms). However, the highly correlated nature of the basis still requires special handling to ensure minimal numerical instabilities (see Appendix \ref{appx:ICs:TechDetails} for more details).

In practice, we use $N = 15$ functions, $q_i$, in the sum of Eq.~\eqref{eqn:kernel:1}, which results in 680 possible independent $Q_a$ after imposing permutation symmetry. 
As mentioned before, we check qualitatively and quantitatively that the final approximation to a given template is sufficient. 
Some templates we consider can be accurately reproduced using fewer modes, while a minority benefit from $N = 25$ (2925 modes). 
Note that $N$ only sets the \textit{highest order} of modes being considered. In practice, we perform a guided sub-sampling---that is explicitly designed to improve numerical stability---as detailed in Appendix \ref{appx:ICs:TechDetails}. So, for example, $N = 15$ does not mean we use 680 mode functions. In practice, we use between 100 to 200 modes depending on the complexity of the target template.

Once we have obtained the coefficients $\alpha_{ijk}$, as defined in Eq.~\eqref{eqn:kernel:1}, we can construct the kernel required to generate a chosen bispectrum. For each term in the shape function, the kernel is written as
\begin{equation}\label{eqn:kernel:2}
    K_{12, ijk} = \alpha_{ijk} \, \frac{g_k(\kThree)}{2P(\kOne)P(\kTwo)} \Big[g_i(\kOne)g_j(\kTwo) + g_j(\kOne)g_i(\kTwo)\Big] + 2 {\rm\, perm.}\,,
\end{equation}
where we have used the notation, $g_i(k) = q_i(k)/k^2$. The $1/k^2$ term arises from converting the shape function back to the bispectrum. Note that in practice, we do not use the permutations for every combination, $ijk$. This is a choice in defining the kernel and still generates the correct bispectrum; see Appendix D of \citetalias{Scoccimarro2012PNGs}. We will, however, utilize the permutations for a small subset of kernels ($i,j,k \leq 3$) for canceling divergent behaviors that we discuss now.

The impact of the kernel is not just on the bispectrum. Specifically, applying the kernel $K_{12} = \sum_{ijk} K_{12,ijk}$ to the initial conditions produces a 1-loop contribution to the primordial power spectrum,
\begin{equation}\label{eqn:ICs:1loop}
    \delta P^{\text{1-loop}}(k) = \int d^3q\, K^2_{12}(q, |\vec{k} - \vec{q}\,|, k) P(q)P(|k -  q|)\,.
\end{equation}
It is crucial to ensure that the kernel does not introduce divergences, causing the one-loop correction to grow more rapidly than the tree-level part, which scales as $P^{\rm tree}(k) \propto k^{n_s - 4}$. 
Otherwise, the introduction of PNG would artificially alter the power spectrum.

Following \citetalias{Scoccimarro2012PNGs}, the preservation of the tree-level power spectrum can be done by ensuring that the combination $g_3(k)^2/P(k)$ is not divergent. In their approach, this was done by inspecting each relevant kernel and using the degrees of freedom from the permutations to cancel terms that were divergent in $\kThree$. 
In general, for the original (not modified) Legendre basis, \textit{all} polynomials would have contributions from some kind of divergent term. 
All Legendre polynomials have a constant term,\footnote{Only even-degree Legendre polynomials have a constant term, but this is for the function argument $\Tilde{k}$. If we rewrite the polynomial in $k$---which is used to define $\Tilde{k}$, as shown in Eq.~\ref{eqn:kbar}---then all polynomials (of both odd and even degree) have constant terms.} which translates to a $1/k^2$ term when we convert from $q(k)$ to $g(k)$. 
We have already resolved this issue via the subtraction of a constant factor when defining our basis functions; see Eq.~\eqref{eqn:modefunc}. 
The modified polynomials have a linear term, which translates to a divergence of $1/k$ in $K_{12}$. 
However, these divergences are subdominant to the tree-level power spectrum, and therefore are not necessary to cancel. 

This point also highlights our choice to write all basis functions in $k$ rather than $\log k$. Most collider templates have terms oscillating in $\log k$ rather than $k$ (Section \ref{sec:sims:Models}), so one may consider it natural to write the Legendre polynomials in $\log k$ instead. However, this choice makes it challenging to avoid infrared (IR) divergences in the one-loop power spectra. Our templates would no longer have terms like $(k + k^2  + \ldots)/k^2$ where the IR scaling is $1/k$ at worst and exhibited by a single term. Instead, we would have terms like $(\log k + \log^2 k  + \ldots)/k^2$, where a majority of the terms effectively scale as $1/k^2$. 
Thus, writing the Legendre polynomials in linear $k$ is a more natural choice for removing divergences in the one-loop power spectrum.

In summary, for our current setup, we do not need to alter the functions $q_{i > 3}$ (beyond the linear subtraction we have already performed) to prevent IR divergences. For the lower-order terms, $q_{i \leq 3}$, we cancel out divergences using the same approach as \citetalias{Scoccimarro2012PNGs}: we use symmetries of the kernels in wavenumbers $\kOne, \kTwo, \kThree$ to construct linear combinations that will still generate the relevant bispectrum but can be combined to cancel any divergent behaviors in the one-loop power spectrum. We first write down the kernel in a slightly different way,
\begin{equation}
    K_{12} = K_{12}^{\rm orig} + K_{12}^{\rm corr}\,,
\end{equation}
where $K_{12}^{\rm corr}$ is the correction term that includes the symmetry-based linear combinations for the lower-order kernels, and $K_{12}^{\rm orig}$ is the fiducial template described previously. In practice, we rewrite all lower-order basis functions ($q_i$ with $i \leq 3$) to explicitly include the allowed permutations over $k_1, k_2, k_3$ and constrain the relative amplitudes/contributions of each permutations such that divergent behaviors are canceled. This follows the process used in \citetalias{Scoccimarro2012PNGs} with some notable extensions. We detail the exact procedure in Appendix \ref{appx:ICs:Divergent}. After this correction, the 1-loop power spectrum of any given kernel is guaranteed to be subdominant to the tree-level prediction, i.e., the 1-loop correction has no terms of the kind $k^{-n}$ with $n > 2$. We explicitly verify this in Appendix \ref{appx:Validation:Pk} for the templates considered in this work. We have also verified that our method here can reproduce (within numerical noise) the results for the standard templates---local, equilateral, and orthogonal---implemented in the 2LPTPNG code.\footnote{\url{https://github.com/dsjamieson/2LPTPNG}} This is an additional confirmation that our non-Gaussian effects are simulated accurately and reliably, as 2LPTPNG is known to produce the expected late-time linear matter bispectrum \citep[][see the large-scale measurement in their Figure 2b]{Coulton2022QuijotePNG}.

Finally, we make two notes of recent work on this topic and highlight differences compared to our formalism presented here. First, \citet{Goldstein:2024:CosmoColl, Goldstein:2025:CosmoColl} have simulated collider-inspired models by considering phenomenological templates that are manifestly separable but still mimic some of the key signatures of the bispectrum from such models. Their work focuses on the squeezed limit of the matter bispectrum, and their simulations are designed to include collider signals from that specific limit. Their approach provides a simple, but effective, method for studying collider signatures in the squeezed limit. 

In contrast, the method introduced in this work is not limited to any specific limit, but is generalized to the full bispectrum across the entire Fourier-space volume. Such an approach is necessary for capturing the complete impact of a cosmological collider model on all aspects of late-time structure formation. 
This is especially necessary when considering arbitrary statistics of the weak lensing (or matter density) field, which take contributions from all limits of the bispectrum and not just the squeezed limit. 
Our basis decomposition also provides a generalized way to study the exact target template, rather that being limited to a specific (squeezed-limit) regime where the template is manifestly in a factorized form.

Second, \citet{Fondi:2025:Gengars} also tackle the generation of initial conditions for arbitrary (separable) bispectrum templates. 
They introduce a modification to the method of~\citetalias{Scoccimarro2012PNGs} that improves its accuracy by further suppressing IR divergences in the one-loop primordial power spectrum.
This is achieved by taking the kernel to be the reduced bispectrum, $K_{12} = B(\kOne, \kTwo, \kThree) / (P(\kOne)P(\kTwo) + \text{perms.})$; see Eq.~(2.11) in \citet{Fondi:2025:Gengars}. 
An important caveat, however, is that their method requires the target bispectrum to be in a manifestly separable form. 

In comparison, our approach directly addresses the challenge of rendering arbitrary bispectra in a separable form.
This was essential because all cosmological collider models, as well as many of the models presented in \citet[][henceforth \citetalias{paper2}]{paper2}, are inherently non-separable.
In addition, we explicitly measure and confirm that the one-loop power spectrum, for all initial conditions we generate, is subdominant to the tree-level term (Appendix \ref{appx:Validation} and Figure \ref{appx:Validation:Pk}) and thus has no tangible impact on any results presented here.

\textbf{Computational runtime:} Under the method we introduce above, generating arbitrary non-Gaussian initial conditions occupies only a fraction (less than 5\%) of the total runtime in producing a single simulation. In particular, for a basis with 680 terms, the initial conditions can be generated in under 15 minutes for $N_{\rm grid} = 1024$. The basis decomposition---which only needs to be done a single time per model and per simulation boxsize (the latter sets the minimum and maximum wavenumbers of the decomposition)---takes under 6 minutes. All timings are estimated with 48 cores from an Intel GOLD 6248R architecture that has maximum core clock-speed of 3 GHz. The initial conditions generation is MPI-enabled (by virtue of being a modification of the MPI-enabled 2LPTPNG code) and can be trivially scaled to even larger grids.

\subsection{Cosmological Collider Models} \label{sec:sims:Models}

Having described the methodology for generating arbitrary non-Gaussian initial conditions, we now consider a variety of collider models to simulate. Our choices of models to consider follow those explored in \citetalias{Sohn:2024:Colliders}. As a reminder, these models represent the impact of some additional (normally heavy) particles on the inflaton field. 
We classify the models into three kinds: (i) the additional particle is a massless/massive scalar, (ii) the particle has nonzero spin, and (iii) the sound speed of the inflaton field or additional field is less than the speed of light. 

In all cases, we normalize the template such that $S(k, k, k) = 1$.\footnote{The one exception, following \citetalias{Sohn:2024:Colliders}, is for the ``equilateral collider,'' and we detail this further below. In many templates (e.g.,~Eq.~\ref{eqn:template:scalar_I}) we do not present the normalized template for simplicity. However, we stress that this normalization is indeed performed when generating our initial conditions.} 
We use the notation $\kappa=k_1k_2k_3/k_T^3$ and $k_T = \kOne + \kTwo + \kThree$ throughout, and define $\mu \equiv \sqrt{m^2/H^2 - 9/4}$ and $\nu \equiv \sqrt{9/4 - m^2/H^2}$, where $m$ is the mass of the particle and $H$ is the Hubble scale during inflation. 
We will denote the inflaton by $\phi$, the additional field by $\sigma$, and spatial and time derivatives as $\partial_i \phi$ and $\dot{\phi}$, respectively. 
All sound speeds ($c_s, c_\sigma$) are in units of the speed of light.

Many templates, such as the scalar-exchange templates and spin-exchange templates below, are often characterized into analytic and non-analytic contributions. The latter are oscillatory behaviors in the bispectrum induced by particle exchange, while the former are simpler polynomials in $k$. Some works focus on specific limits of a bispectrum (\eg, the squeezed limit) where these two contributions can be represented as additive terms and analyzed separately. The templates available to our work are designed for the entire bispectrum domain, rather than a specific limit, and do not admit such a simple additive representation. As a result, we are unable to show the impact of just the analytic or just the non-analytic contributions on non-linear structure. This is purely a limitation from the provided theoretical templates; our basis-decomposition method will work for both analytic and non-analytic contributions.

\begin{figure}
    \centering
    \includegraphics[width=1\columnwidth, trim = 2cm 9cm 2cm 0cm, clip]{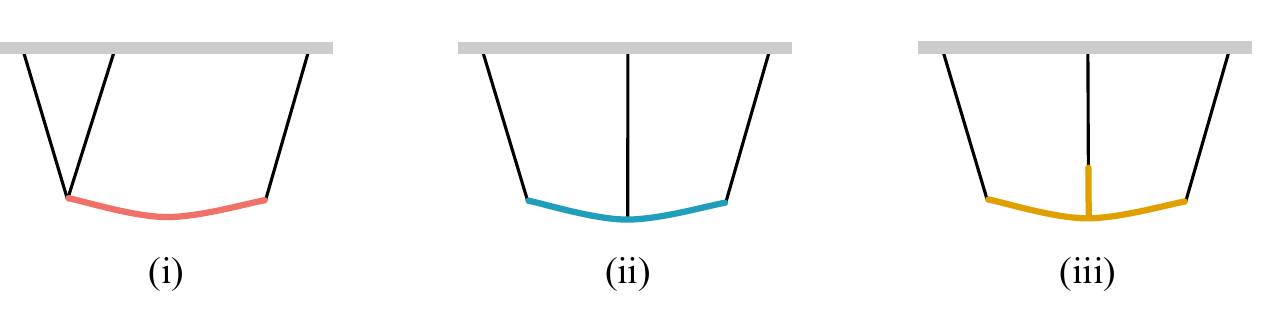}
    \caption{Feynman diagrams for the classes of interactions considered in this work. The black (colored) lines represent inflaton fluctuations (massive particles). Panels (i), (ii), and (iii) correspond to single-, double-, and triple-exchange of massive particles, respectively. Case (i) includes the SI, SII, HSC, and EC models; (ii) represents the MSC scenario; and (iii) corresponds to the QSF scenario.}
    \label{fig:Feynman}
\end{figure}

\subsubsection{Scalar Exchange}

We begin with scalar-exchange diagrams, illustrated schematically in Figure~\ref{fig:Feynman}, where inflaton fluctuations interact through the exchange of a massive scalar field. In what follows, we consider three representative scenarios of scalar exchange.

\paragraph{Quasi-Single Field (QSF) Inflation} 
This model captures the impact of massive (but still relatively light) particles with $m/H \leq 3/2$. This shape is generated by a cubic interaction, $\sigma^3$, in the heavy sector together with linear mixing, $\dot\phi\sigma$. We use the standard template defined by~\citep{Chen:2010:QSF}
\begin{mybox}    
\vskip 10pt
\begin{equation} \label{eqn:template:qsf}
    S^{\rm QSF} = 3\sqrt{3\kappa} \,\frac{Y_\nu(8\kappa)}{Y_{\nu}({8}/{27})}\,,   
\end{equation}
\end{mybox}
\noindent where $Y_\nu$ is the Bessel function of the second kind. While there are recommended techniques for factorizing this particular template (see, e.g.,~\citet{Dizgah2020CosmoCollider}), we instead factorize the template using our systematic methodology discussed above.

\paragraph{Scalar I} Another class of exchange diagrams involves a single exchange of a heavy scalar, arising from a cubic interaction of the form $(\partial\phi)^2\sigma$. 
Due to the breaking of time translation symmetry during inflation, we can separately consider two types of interactions $\dot{\phi}^2\sigma$ and $(\partial_i\phi)^2\sigma$. We will refer to these Scalar I and II cases, respectively.
The bispectra generated by these interactions have been computed using cosmological bootstrap techniques~\citep{Arkani-Hamed:2018kmz,Baumann:2019oyu,Pimentel:2022fsc,Jazayeri:2022kjy}, and 
\citetalias{Sohn:2024:Colliders} provide the following template that approximates the exact shape:
\begin{mybox}
{\small
\begin{align} \label{eqn:template:scalar_I}
    S_{\rm col.}^{\rm SI} & = \frac{2k_1k_2k_3}{\beta k_T^2k_{12}}\bigg[ 1+ \frac{4k_3}{ (\beta+2) k_{12}} + \frac{(\beta+4)k_3^2}{ (\beta+2)^2 k_{12}^2}\bigg] \bigg(\frac{k_T}{k_{12}}\bigg)^{-\frac{\beta}{\beta+2}}\nonumber\\
    & - \frac{k_1k_2k_3}{6\cosh(\pi\mu)k_{12}^3}\bigg[{2(2\mu^4-1)}\bigg( \frac{k_3^2}{k_T^2}+ \frac{k_3(4k_T-k_3)}{k_T^2} \log \bigg(\frac{k_T}{k_{12}}\bigg)  +\log^2\bigg(\frac{k_T}{k_{12}}\bigg) \bigg) \nonumber\\
    & - \mu^2 \bigg(\frac{k_3}{k_{12}}\bigg)^{\frac{1+16\mu^2}{1+8\mu^2}} \bigg(\frac{k_3(6k_T-k_3+8\mu^2(8k_T-k_3))}{(1+8\mu^2)k_T^2} + \frac{2(3+68\mu^2+384\mu^4)}{(1+8\mu^2)^2}\log \bigg(\frac{k_T}{k_{12}}\bigg)
    \bigg)\bigg]\nonumber\\
    & + \frac{k_1k_2}{k_{12}^2} \sqrt{\frac{\pi^3\beta(\beta+2)}{\mu \sinh(2\pi \mu) }} \bigg(\frac{k_3}{k_{12}}\bigg)^{\frac{1}{2}} \cos\bigg[\mu\log\bigg(\frac{k_3}{2k_{12}}\bigg) + \delta \bigg] + 2~{\rm perm.} \,,  
\end{align}
}
\end{mybox}
where $k_{ij}= k_i+k_j$, $\beta = \mu^2 + 1/4$, and $\delta = \arg [\Gamma({\frac{5}{2}}+i\mu)\Gamma(-i\mu) (1+i\sinh\pi\mu)]$.
This template consists of a non-oscillating piece and an oscillating term. 
The latter is of comparable amplitude to the former for low masses ($m/H \lesssim 3/2$) but becomes exponentially suppressed for high masses ($m/H \gg 3/2$).
The oscillatory frequency is fixed by the mass of the particle in Hubble units.
In the limit of $m/H \gg 3/2$, the above template approaches the equilateral shape.

\paragraph{Scalar II} This cases parallels the Scalar I case, but for the interaction $(\partial_i \phi)^2 \sigma$. 
In practice, the actual Scalar II template we use is defined as a linear combination of two interactions 
\begin{equation} \label{eqn:template:scalar_II}
    S_{\rm col.}^{\rm SII} = - S_{\rm col.}^{\rm SI} - \frac{14}{100} S_{\rm col.}^{\rm sp}\,,
\end{equation}
where $S_{\rm col.}^{\rm SI}$ is given in Eq.~\eqref{eqn:template:scalar_I} and the second term ($S_{\rm col.}^{\rm sp}$), corresponding to the actual interaction $(\partial_i \phi)^2 \sigma$, is given as in \citet{Sohn:2024:Colliders}. 
The expression is quite lengthy, and so we quote it in Eq.~\eqref{eqn:template:SII} in the appendix. 

The motivation for this combination stems from the fact that the bispectra generated by interactions $\dot{\phi}^2\sigma$ and $(\partial_i \phi)^2 \sigma$ are highly correlated. To construct a distinguishable template, we define linear combination that is orthogonal to the Scalar I shape. 
This approach follows the definition of the standard orthogonal template, which itself is a linear combination of templates from two cubic interactions in single-field inflation \citep{Senatore2010WMAP5pngs}. In the limit $m/H \gg 1$, the resulting Scalar II shape closely resembles this orthogonal template. 

\subsubsection{Spin Exchange}\label{sec:sims:Models:Spin}

If the inflaton has interactions with particles of spin $s>0$, then the resulting bispectrum gains a characteristic angular dependence. 
In the limit of $m \gg H$, this intermediate field, $\sigma$, can be integrated out, resulting in a single-field effective theory of the inflaton; we will call this scenario ``heavy spin collider'' (HSC). 
In this case, the exact shape is given by~\citep{MoradinezhadDizgah:2018ssw}
\begin{mybox}
\begin{align}
     S^{\rm HSC} & = \mathcal{P}_s(\hat{k}_1\cdot \hat{k}_3) \frac{k_2(k_1k_3)^{s-1}}{ k_T^{2s+1}}\bigg[ (2s-1)( k_{13}k_T+2s k_1 k_3) +k_T^2 \bigg] + {\rm 5~ perm.}\,,\label{Sspin}
\end{align}
\end{mybox}
\noindent where $\mathcal{P}_s$ is the degree-$s$ Legendre polynomial of the angle $\hat{k}_1\cdot \hat{k}_3 \equiv (\kOne^2 + \kThree^2 - \kTwo^2)/(2\kOne\kThree)$.
This template is largely similar to the equilateral shape, but with an angular dependence.
As emphasized in \cite{MoradinezhadDizgah:2018ssw}, its amplitude peaks away from the equilateral limit so the normalization convention of $S(k,k,k) = 1$ increases the overall amplitude of this bispectrum relative to other shapes.
This highlights the somewhat subtle choices in normalizing a template and the care needed when interpreting $\fNL$ constraints across different shapes.

\subsubsection{Nontrivial Sound Speed}

In the EFT approach to inflation~\citep{Cheung:2007st}, boost symmetries can be broken and the sound speed of perturbations can be much smaller than the speed of light. 
It is also possible for the inflaton and the intermediary particle to have significantly different sound speeds, which also enhances the amplitude of the non-Gaussianity, normally in the squeezed limit. 
We follow \citetalias{Sohn:2024:Colliders} in probing a few phenomenological templates of this type.

\paragraph{Equilateral Collider} If we allow the sound speed to vary, the interaction $\dot{\phi}^2\sigma$ now results in a template of the form,
\begin{mybox}
\begin{align}
    S^{\rm EC} = \,\,& \frac{k_1k_2}{k_{12}^2} \bigg(\frac{k_3}{k_{12}}\bigg)^{\frac{1}{2}}\cos\bigg[\mu\log\bigg(\frac{c_\sigma k_3}{2c_sk_{12}}\bigg)+\delta^{\rm EC} \bigg]  + {\rm 2~perm.} 
\end{align}
\end{mybox}
Note that we can subsume the sound speed factors into the phase as $\phi^{\rm EC} = \mu\log(c_\sigma/c_\sigma) + \delta^{\rm EC}$. We will vary the mass and this effective phase for this template. Note that the normalization of this template is somewhat different, as we normalize such that $S^{\rm EC}(k, k, k) = \cos[\mu\log(\frac{c_\sigma k_3}{2c_sk_{12}})+\delta^{\rm eq.col.}]$. This avoids large changes in the normalization arising from fine-tuned cancellations enabled by the free phase parameters, since $\cos(\cdots) = 0$ for certain choice of parameters.

\paragraph{Low-Speed Collider} In the limit that $c_s \ll c_\sigma$, i.e., the sound speed of the inflaton perturbations\footnote{In this regime, the standard single-field inflaton description may break down. 
A more appropriate framework is the EFT of inflationary perturbations~\citep{Cheung:2007st}, where the relevant degree of freedom is the Goldstone boson of broken time translations. 
For simplicity, we will continue to refer to this degree of freedom as the ``inflaton,'' denoted by $\phi$.} is significantly slower than that of the massive field's, we can derive the template \citepalias{Sohn:2024:Colliders}
\begin{mybox}
\begin{align}
    S^{{\rm LSC}} =  S^{\rm eq} +
 \frac{k_1^2}{3k_2k_3} \bigg[ 1+ \alpha \bigg(\frac{k_1^2}{3k_2k_3}\bigg)^2\bigg]^{-1} + {\rm 2~perm.}\,,
\end{align}
\end{mybox}
where $\alpha \equiv c_s m/H$. For $\alpha = 0$ ($\alpha = 1$) the template is correlated with the local (equilateral) shape. At intermediate values it is correlated somewhat with the orthogonal shape. Lower values of $\alpha$ result in the bispectrum peaking at larger $k$ in the squeezed limit.

\paragraph{Multi-Speed Collider} Considering a massless additional field with higher derivative interactions, such as $\dot{\phi}\dot{\sigma}$ or $\partial^2_i \phi \sigma$, can lead to the inflaton field inheriting the sound speed of the intermediate field. Furthermore, each leg of the three-point interaction would have its own sound speed. The shape template can then be written down as
\begin{mybox}
\begin{align}
    {S}^{\rm MSC}= \frac{k_1k_2k_3}{(c_1 k_1+ c_2 k_2 +c_3 k_3)^3} +  5~{\rm perm.}
\end{align}
\end{mybox}
In our analysis, we fix $c_3 = 1$ with the condition $c_1 \leq c_2 \leq c_3$.

\begin{figure}
    \centering
    \includegraphics[width=\columnwidth]{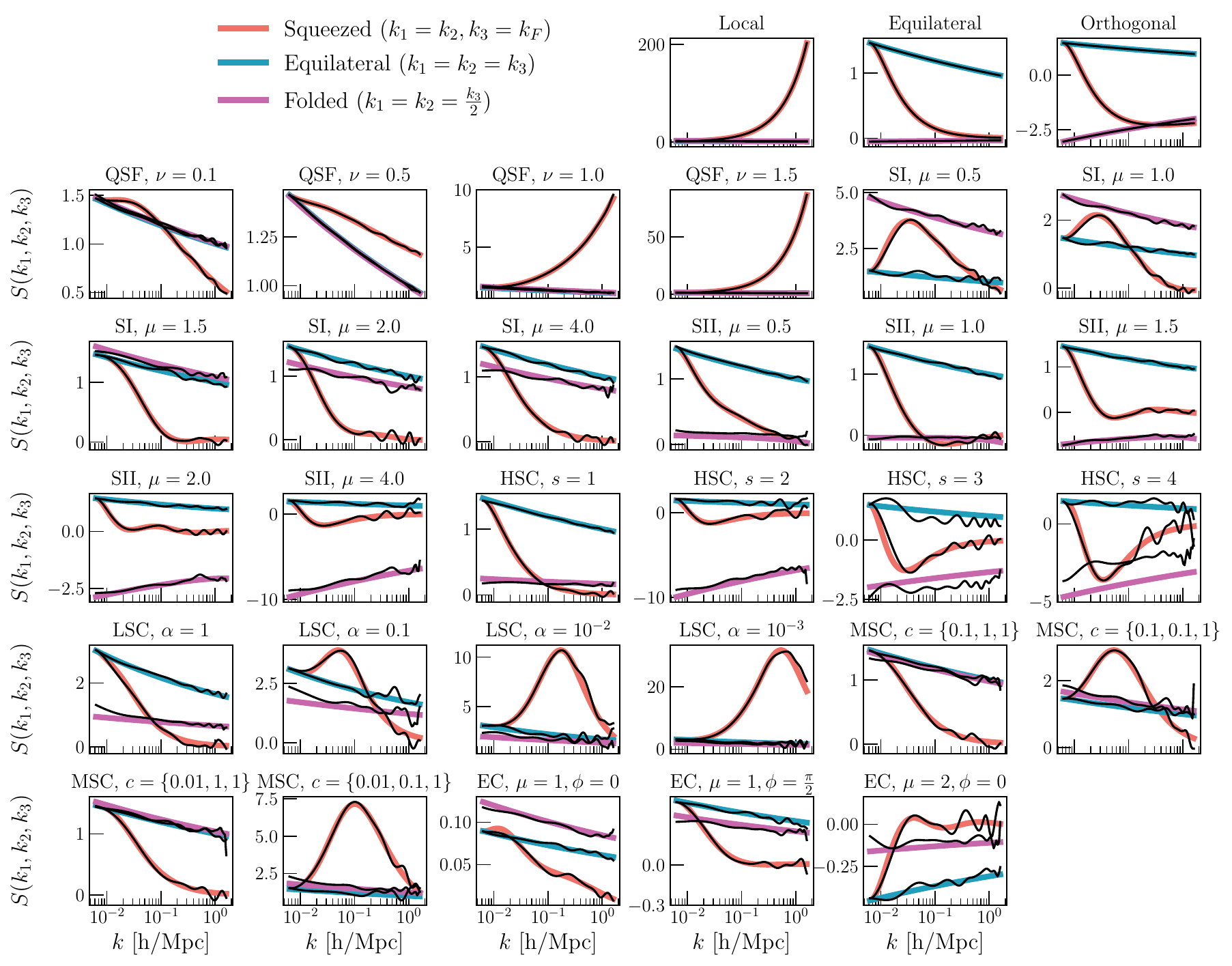}
    \caption{The different templates considered in this work, shown in three specific limits, alongside the approximated versions using our basis functions (black lines). In all cases, the templates are adequately reproduced by our approximations. There are some oscillatory residuals in the approximated template, given our basis functions are Legendre polynomials in linear $k$. These can be suppressed by adding more modes and allowing for fine cancellations between basis functions, but such fine-tuning degrades the numerical stability of our basis set (see Section \ref{sec:sims:ICs} and Appendix \ref{appx:ICs:TechDetails}). For this reason, we retain some oscillatory deviations in the approximated template but note that in all cases the residuals are smaller than the overall variation of the bispectrum (in a given limit) across $k$.}
    \label{fig:Template}
\end{figure}

\subsection{Shape Validation}

Figure~\ref{fig:Template} shows the bispectrum templates considered in this work (colored lines), alongside their approximated versions constructed from our basis functions (black lines). We show the squeezed, equilateral, and folded limit of the bispectra. In all cases, the approximated templates provide a good match to the true templates. 

Some deviations are visible, particularly as small oscillations at high $k$. 
These arise because our basis functions consist of Legendre polynomials, which are inherently oscillatory. 
In regimes where the template is smooth, the functions $Q_b$ require fine cancellations to recover the smooth, non-oscillatory behavior. 
This can lead to amplifications in the condition number of the basis matrix (see Appendix~\ref{appx:ICs:TechDetails}) and can induce numerical instability. 
To avoid this, we allow mild oscillatory residuals at the benefit of improved numerical stability.
Importantly, these residuals are still smaller than the overall template/variations across the Fourier-space volume. 
The largest discrepancy occurs for the HSC model with $s = 4$, where the approximation underestimates the amplitude in the folded limit by $\sim 30\%$. 
Nonetheless, we still simulate this template as it accurately captures the squeezed and equilateral limits, and the error in the folded limit only \textit{reduces} the amplitude of the signal from the template. So the error caused by the approximation will not artificially boost the signal of the template.

\begin{figure}
    \centering
    \includegraphics[width=1\columnwidth]{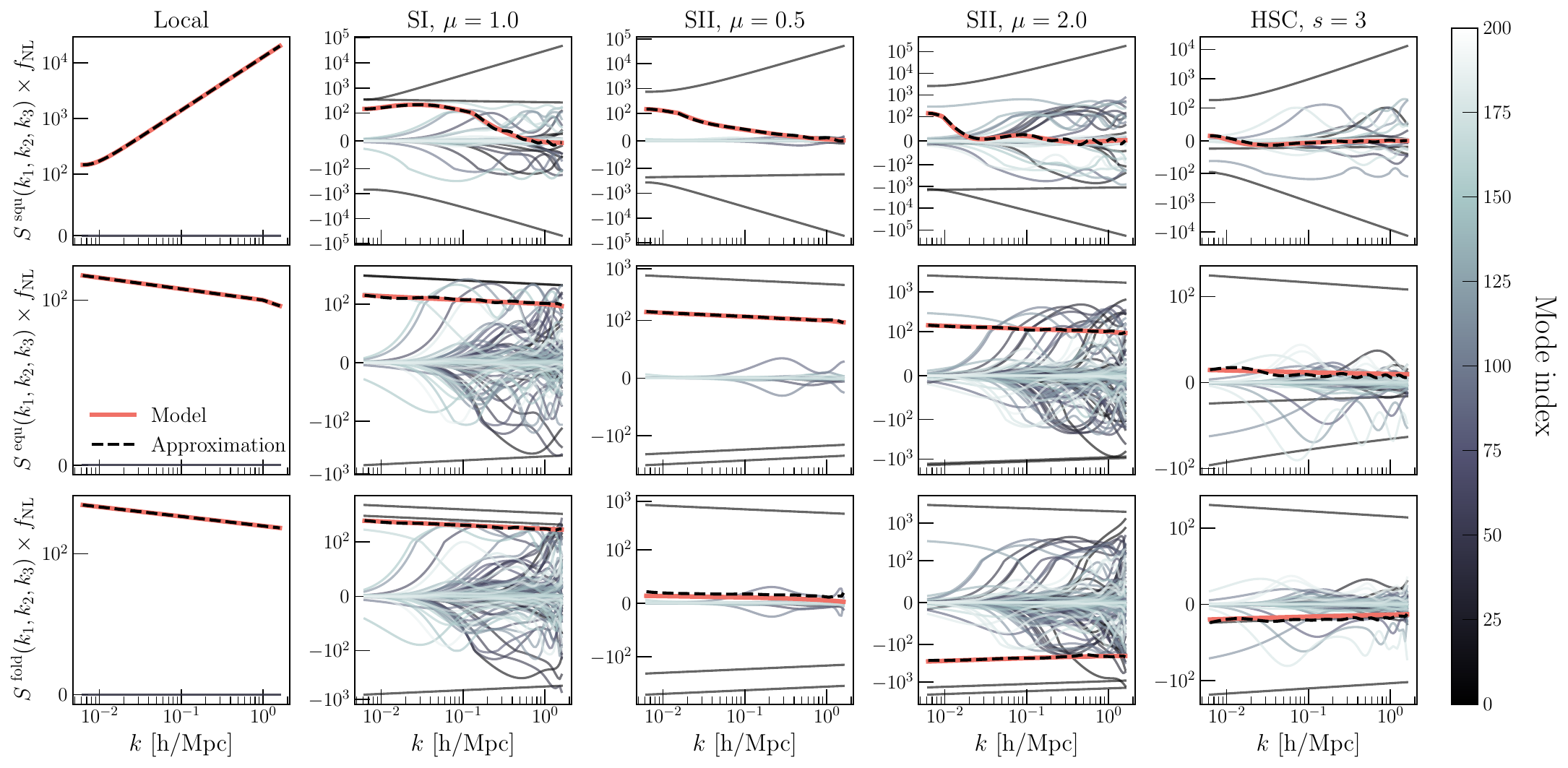}
    \caption{An example of the mode decomposition for a few templates shown in Figure~\ref{fig:Template}. The rows show the shape function, $S$, in the squeezed, equilateral, and folded limits. The columns show different models, following the same nomenclature as Figure~\ref{fig:Template}. The analytic model (red) is shown alongside the approximation (dashed black line). The individual modes are shown in the thin, translucent lines and are colored by the mode index. Larger indices generally correspond to higher-order modes. The local template is reproduced exactly using just a single mode (hidden behind the red line). The rest of the templates require some cancellations between different basis functions in order to reproduce the target template. We use a symmetric log axis to highlight the presence of a few large-amplitude modes that are canceled somewhat finely to construct the target template.}
    \label{fig:Decomposition}
\end{figure}

Figure~\ref{fig:Decomposition} provides a brief visualization of how the different basis modes combine to reproduce a given template. 
We show five different templates in the squeezed, equilateral, and folded limits. 
We see that the collider templates require significant cancellations in the oscillatory basis functions in order to approximate the targeted, \textit{smooth} template. 
Appendix~\ref{appx:ICs} describes in detail our numerical approach to handling the stability of such cancellations in our method.

When generating simulations, we must decide what value of $\fNL$ to use to inject a bispectrum signal in the initial conditions. Using a larger value helps the signal emerge above the numerical noise and provides a more precise characterization of the signal. However, PNGs are perturbative expansions of the (Gaussian) initial conditions and we are currently only using the tree-level (zeroth-order) estimate of the bispectrum. If we use too large a value of $\fNL$, then it will no longer suffice to consider just the tree-level calculation as the first-order will become more relevant. This is briefly discussed in Appendix C of \citetalias{Scoccimarro2012PNGs}. Balancing these considerations, we use $\fNL = 100$ as our default value, with two exceptions: $\fNL = 10$ for the models with nonzero spin and deviations in the sound speed, and $\fNL = 200$ for the equilateral collider. The variable choice accounts for the variable impact of normalization on the different templates.\footnote{For example, if a template has minimal power in the equilateral limit relative to the squeezed/folded limits then normalizing it to $S(k, k, k) = 1$ causes the template to produce significantly more power for a value of $\fNL = 1$, relative to other templates such as the equilateral or orthogonal types. This is also discussed in Section 3.1 of \citet{Sohn:2024:Colliders}.}

\section{Cosmological Collider Signals in the Nonlinear Universe}\label{sec:results}

We now present the signatures of collider models in various aspects of nonlinear structure formation. We start in Section~\ref{sec:results:lensing} by quantifying the Fisher information in lensing observations from upcoming datasets. We then examine their impacts on the matter field in Section \ref{sec:results:matter} and on the halo field in Section \ref{sec:results:halos}.

\subsection{Information in Weak Lensing Observations}\label{sec:results:lensing}

Weak lensing has traditionally not been considered a probe of PNGs. However, the availability of $N$-body simulations with PNG signals now allows it to become an accessible probe for such analysis and to provide its unique advantages for such constraints, the primary of which are that it is a direct (unbiased) tracer of the density field, it is easier to analyze with simulation-based models, and it provides constraints on much smaller scales ($k \approx 1 \hinvmpc$) than existing analyses. Furthermore, cross-correlations of lensing and galaxy fields can self-calibrate galaxy bias parameters that are vital for analyses of PNGs with galaxy clustering. These aspects are discussed in more detail in Section 5.1 of \citet{Anbajagane2023Inflation}.

We now discuss the forecasted constraints on the different PNGs using weak lensing measurements from the upcoming LSST Y10 dataset \citep{LSST2018SRD}. The full forward model mimics that of \citet{Anbajagane2023CDFs, Anbajagane2023Inflation} and its description is reproduced briefly in Appendix \ref{appx:ForwardModel}. We assume a footprint of 14,000 $\deg^2$ with a source density of $n_{\rm gal} = 30\,\, {\rm arcmin}^{-2}$, following the LSST Y10 prescription. We use the same LSST redshift distributions assumed in \citet{Anbajagane2023Inflation}, which is a more conservative version of that described in \citet{LSST2018SRD}. Our primary observable is the lensing convergence field, $\kappa$, which is a line-of-sight integral of the density field and is defined by
\begin{equation}\label{eqn:convergence_definition}
    \kappa(\nhat, z_s) = \frac{3}{2}\frac{H_0^2\Omega_{\rm m}}{c^2}\int_0^{z_s}\delta(\nhat, z_j) \frac{\chi_j(\chi_s - \chi_j)}{a(z_j)\chi_s}dz_j\frac{d\chi}{dz}\bigg|_{z_j},
\end{equation}
where $z_s$ is the redshift of the ``source'' plane/galaxies whose light is being lensed, $\nhat$ is the line-of-sight direction on the sky, $\delta$ is the overdensity field, $\chi$ is the comoving distance from an observer to a given redshift, $a$ is the scale factor, $H_0$ is the Hubble constant, $\Omega_{\rm m}$ is the matter energy density fraction at $z = 0$, and $c$ is the speed of light. We have used the shorthand $\chi(z_s) \equiv \chi_s$ and $\chi(z_j) \equiv \chi_j$.

We use the second and third moments of the field as our main summary statistics. These are computed as 
\begin{equation}\label{eqn:Moments}
    \langle \kappa^{(1)}\kappa^{(2)}\ldots \kappa^{(N)}\rangle(\theta) = \frac{1}{N_{\rm pix} - 1}\sum_{i=1}^{N_{\rm pix}}\kappa^{(1)}_i\kappa^{(2)}_i\ldots \kappa^{(N)}_i\,,
\end{equation}
where $\kappa^{(j)}$ is the convergence field of the $j^{\rm th}$ tomographic bin, $N_{\rm pix}$ is the number of pixels in the survey footprint. In all cases,  $\langle\kappa^{(j)}\rangle = 0$ is enforced and the scale dependence on $\theta$ is obtained by making measurements after smoothing the fields with a harmonic-space tophat filter,
\begin{equation}\label{eqn:tophat}
    W_\ell(\theta) = 2\frac{j_1(\ell \theta)}{\ell\theta}\,,
\end{equation}
where $\theta$ is the tophat radius in radians. We use ten bins of $\theta$, between $3.2\arcmin < \theta < 200\arcmin$, which follows the analysis choices of previous work \citep{Gatti2022MomentsDESY3}, and all fields are smoothed by the same $\theta$. As mentioned above, we only consider moments of the order $N = 2$ and $N = 3$. \citet[][see their Figure 6]{Anbajagane2023CDFs} measure a set of moments on DES Y3 data and find the fourth moment is only weakly detected, while the 5th moment is consistent with no cosmological signal. There are also a number of additional possibilities for the choice of summary statistics. The second and third moments of the field have been validated extensively as a robust summary statistic \citep{Gatti2020Moments} and have been used on weak lensing data to constrain cosmology parameters \citep{Gatti2022MomentsDESY3}. Other statistics (which are not considered in this work) have also recently undergone the same validation/treatment \citep[\eg][]{Gatti:2024:WPH, Prat:2025:Homology}. As a result, the lensing-only forecasts we present below are underestimating the complete set of information on PNGs that we can reliably extract from the lensing field.

We estimate the Fisher information matrix as
\begin{equation}\label{eqn:Fisher}
    \textbf{F}_{ij} = \sum_{m,n}\frac{d\widetilde{X}_m}{d\theta_i}\big(\mathcal{C}^{-1}\big)_{mn}\frac{d\widetilde{X}_n}{d\theta_j}\,,
\end{equation}
where $\frac{d\widetilde{X}_m}{d\theta_i}$ is the mean derivative of point $m$ in data-vector $X$ with respect to parameter $\theta_i$. Next, $\mathcal{C}^{-1}$ is the inverse of the numerically estimated covariance matrix and is computed while accounting for the Kaufman-Hartlap factor \citep{Kaufman1967, Hartlap2007},
\begin{equation}\label{eqn:invertcov}
    \mathcal{C}^{-1} \rightarrow \frac{N_{\rm sims} - N_{\rm data} - 2}{N_{\rm sims} - 1} \,\mathcal{C}^{-1}\,.
\end{equation}
The Kaufman-Hartlap factor is $\gtrsim 0.95$ for the datavector used in this work. The Fisher constraints are obtained by inverting the matrix $\textbf{F}$.

In this work, we estimate the mean derivative $\frac{d\widetilde{X}_m}{d\theta_i}$ using 10 simulations per model, which results in 30 independent survey realizations as three independent LSST Y10 footprints can fit within one full-sky mock. This is markedly fewer realizations per model than that produced in \citet{Anbajagane2023Inflation}, and is due to computation and storage limitations, as we are considering eight times as many models in this work. Noisy estimates of the mean derivative can cause artificially tighter constraints by breaking degeneracies between parameters in the Fisher information matrix \citep[\eg][]{Coulton:2023:FisherTest}. However, this effect can be alleviated if we limit ourselves to single parameter analyses where there is no secondary parameter to form a degeneracy in the first place. As a result, we only estimate the Fisher information for the PNG amplitude parameter, $\fNL$. For this setup, we have verified that our Fisher-derived uncertainties on $\fNL$ change by less than 2--3\% for all considered models if we use half as many realizations as our fiducial case. The good numerical convergence is partially because we estimate derivatives using pairs of simulations that share the same random seed but are run with $\fNL \pm X$, where the choice for $X$ is described in Section \ref{sec:sims:Models}. We also note that \citet{Avila:2023:PairFixed} show, in a different context, that such matching improves the precision of statistics derived from the simulation pairs. 

\begin{figure}
    \centering
    \includegraphics[width=0.32\columnwidth]{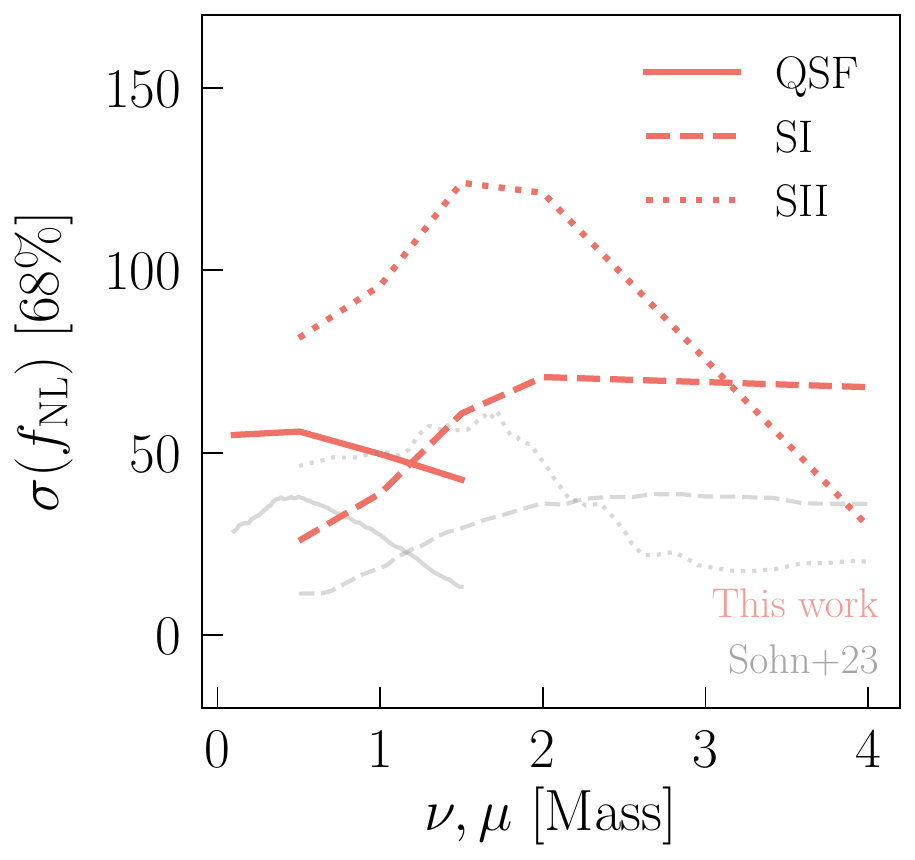}
    \includegraphics[width=0.31\columnwidth]{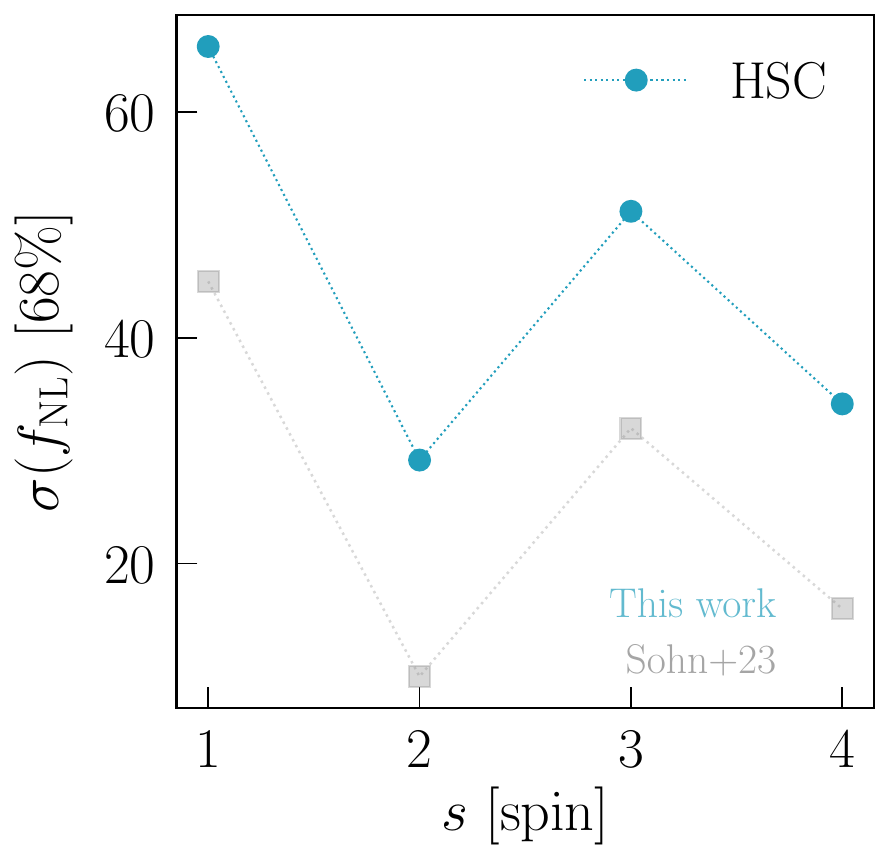}
    \includegraphics[width=0.32\columnwidth]{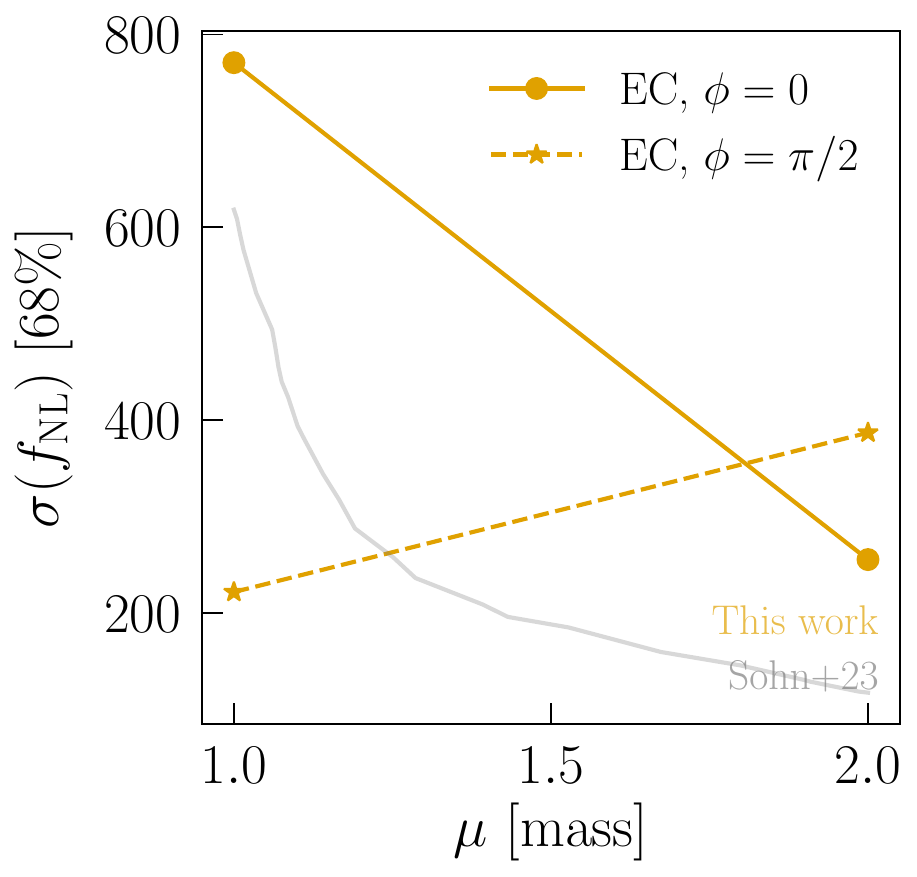}\vspace{10pt}
    \includegraphics[width=0.32\columnwidth]{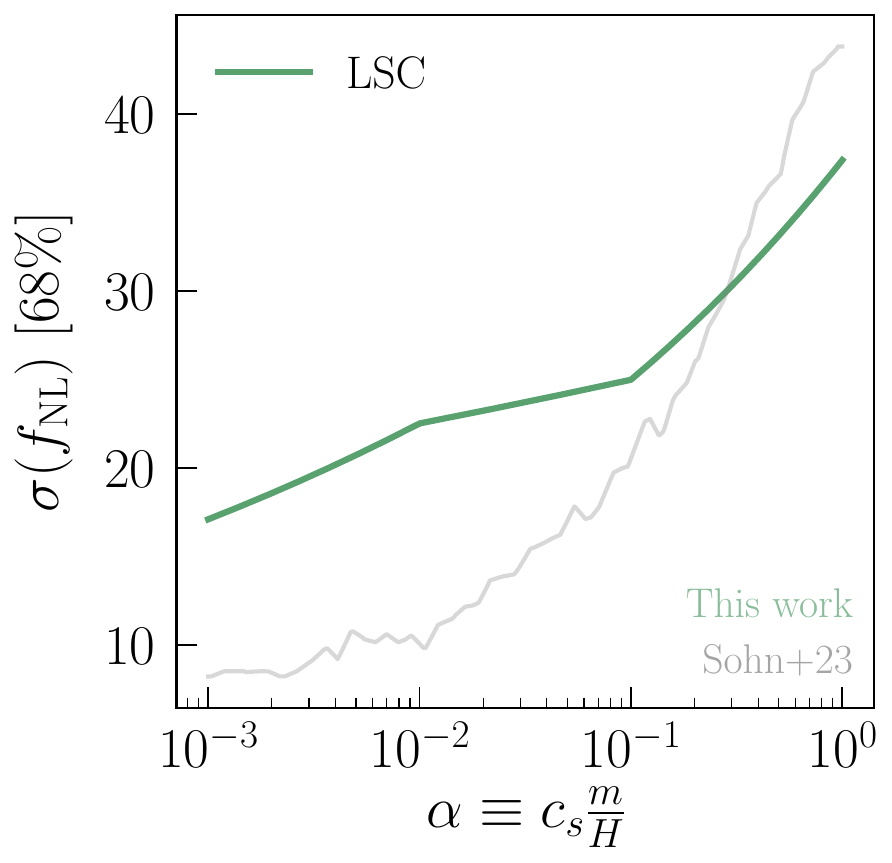}
    \includegraphics[width=0.32\columnwidth]{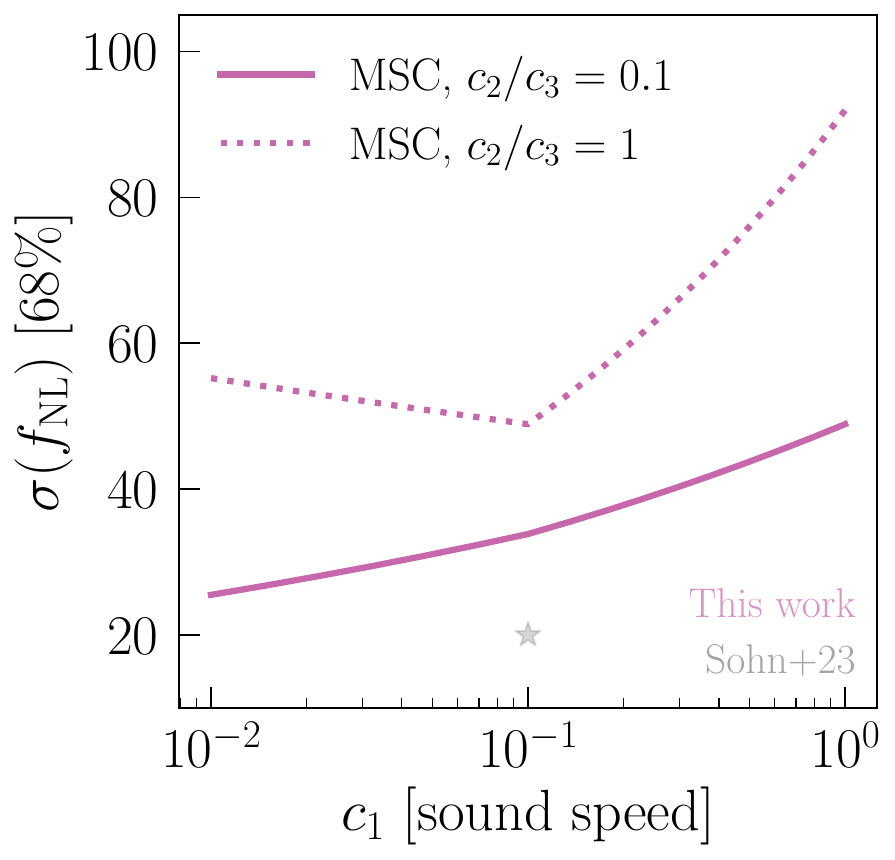}
    \caption{Constraints on the amplitude $\fNL$ for different models/templates and parameter spaces. We compare our forecasted constraints for LSST Y10 lensing against the CMB-derived constraints from \citet{Sohn:2024:Colliders}. The former are within 30\% of the latter for most models studied here, highlighting the primordial information in the lensing field. For the Equilateral Collider (EC), \citetalias{Sohn:2024:Colliders} present the constraints at the maximum likelihood point for the phase $\phi$, whereas we show our forecasts for two choices of $\phi$. This template is also normalized differently (relative  to the other templates), which results in higher values of $\fNL$ (see Section \ref{sec:sims:Models}). For the Multi-Speed Collider (MSC), we can only show CMB constraints at a single point that overlaps with our exact setup, and this is indicated as a gray star. The lensing constraints do not marginalize over additional parameters, where such marginalization is expected to degrade constraints by 20-30\%. See Section \ref{sec:results:lensing} for details.}
    \label{fig:Constraints}
\end{figure}

Figure \ref{fig:Constraints} now presents the forecasts on various collider models for LSST Y10 lensing data. Each model is characterized by the amplitude $\fNL$ as well as 1 or 2 additional parameters such as the mass, spin, and speed, etc. We produce simulations for three to four choices of these additional parameters, with the number of choices depending on the exact model. We can then compute the expected constraint on $\fNL$ given a specific value for these additional parameters. 

Each panel of Figure~\ref{fig:Constraints} showcases one class of models discussed in this work. We overplot the constraints from \citet{Sohn:2024:Colliders} for each setup.\footnote{These constraints are derived from current \textit{Planck} 2018 temperature and polarization data. The upcoming Simons Observatory data, in combination with \textit{Planck} 2018, will is expected to improve constraints by 20--50\% for the local, equilateral, and orthogonal templates \citep{SO:2019:Forecast}.} For the Equilateral Collider (EC) we only produced simulations at two different values for the mass, $\mu$, and so can only extract $\fNL$ constraints at those two values. For the Multi-Speed Collider (MSC), we can only show constraints at a single point that overlaps with our exact setup, and this is indicated as a gray star. These plots also provide an indirect validation of our method, in that the scaling of our $\fNL$ constraining power with a given model parameter (mass, spin, sound speed etc.) broadly matches that seen in the CMB analysis of \citet{Sohn:2024:Colliders}. A key result of Figure~\ref{fig:Constraints} is that the lensing-only forecasts are within roughly 30\% of the CMB-derived constraints from \textit{Planck} 2018. 

As a result of our choice to perform single-parameter constraints, the constraints can be viewed as a best-case scenario when using the second and third moments. We note, however, that these constraints can still be improved upon further if we used additional statistics. \citet{Anbajagane2023Inflation} find that marginalizing over additional parameters---such as $\sigma_8$, the root-mean squared amplitude of fluctuations at $z = 0$ on scales of $8 \mpc/h$, and $\Omega_m$, as well as two nuisance parameters related to the intrinsic alignment of galaxies \citep{Troxel2015IAReview}---degraded constraints on $\fNL$ by 20\% to 30\% for the standard $\fNL$ templates (local, equilateral, and orthogonal). We expect a similar degradation here for the templates we study in this work, given they are somewhat correlated with one or more of the standard templates studied in \citet{Anbajagane2023Inflation}. In summary, the constraints presented below do not marginalize over additional cosmology and IA parameters due to limitations in the size of the simulation suite, but prior work indicates that even after such marginalization, the constraints from LSST Y10 lensing will be competitive with the CMB-derived constraints of \citet{Sohn:2024:Colliders}.

\begin{figure}
    \centering
    \includegraphics[width=\columnwidth]{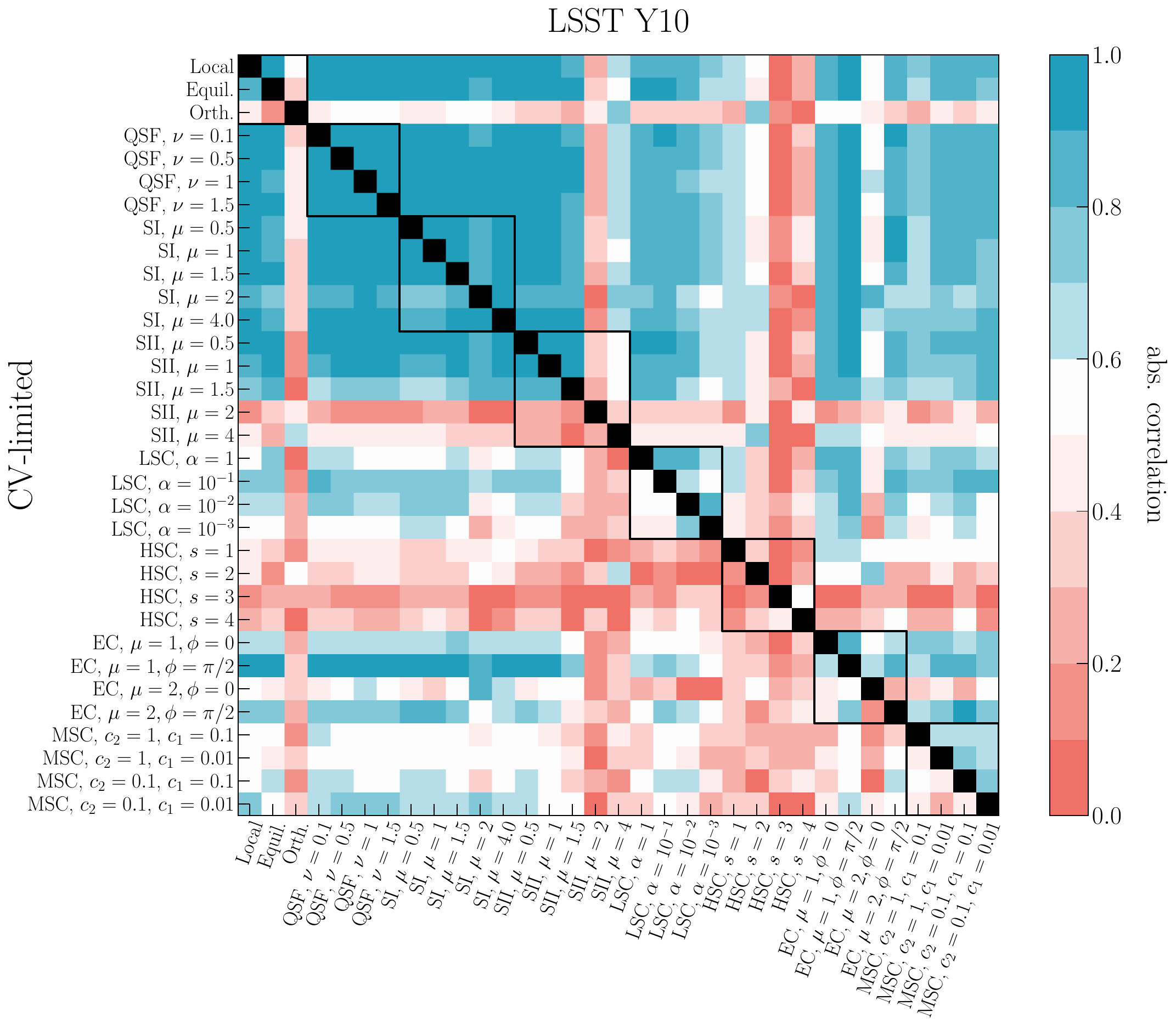}
    \caption{The correlation in the Fisher posterior for pairs of models. Values close to 1 and 0 indicate strong degeneracy and orthogonality, respectively, between the late-time predictions of two models. The upper (lower) triangle shows results for an LSST Y10 survey (cosmic variance, or CV, limited survey). The correlations in LSST Y10 are similar to those from a CV-limited survey. The templates from scalar exchange (QSF, SI, SII) are highly correlated with each other, while those from particle spins and sound speed variations are more orthogonal to each other and to these scalar templates. The diagonal is shaded black for visualization. The black boxes denote different classes of templates.}
    \label{fig:Corr}
\end{figure}

Finally, Figure~\ref{fig:Corr} presents the cross-correlation the late-time signatures of different templates, as seen in the second and third moments of the lensing convergence field. This is computed by estimating the Fisher information a two-model analysis,\footnote{Deviating away from a single-parameter analysis causes our estimates to be impacted by noise in the derivatives. We have checked that our estimates of the off-diagonal covariance changes by 20-30\% if we use half as many realizations. This does not change our qualitative discussions about the correlation between two templates. As a result, our analysis cannot provide any precise constraints on the correlation between two models' late-time signals.} and computing the parameter covariance matrix. We normalize the off-diagonal component to obtain the correlation. We also take the absolute value so as we are only interested in the amplitude (and not sign) of the correlation. Values close to 1 indicate strong correlations between the late-time predictions of two templates, and those close to 0 indicate strong orthogonality between the same. The upper triangle are results for datavectors in an LSST Y10 survey, while the lower triangle is the same for datavectors in a cosmic variance-limited survey, i.e., one with no shape noise in the lensing estimates. The black boxes approximately delineate different classes of the templates discussed in Section \ref{sec:sims:Models}.

We find the late-time predictions of all scalar-exchange templates (QSF, SI, SII) are strongly correlated with each other, and with some of the standard PNG templates (local and equilateral). The templates that vary sound speeds, or consider particles with nonzero spin, are quite uncorrelated with each other and with the scalar-exchange templates. These results highlight our ability (or lack thereof) to constrain the other physical parameters of the model in question, i.e., if we find $\fNL \neq 0$, how reliably can we predict the parameters of the template contributing to that signal? Figure \ref{fig:Corr} shows that for some models (such as the scalar-exchange) this will be challenging, while for others (like the heavy spinning collider, HSC) it may be achievable. This result also highlights that the main benefit of using the exact bispectrum template---as opposed to some approximate, ``catch all'' template---is maximizing the signal-to-noise of the final inference. For example, in the case of the scalar exchange templates, there is a strong correlation with the local template (which is also visually apparent in Figure \ref{fig:Template}). Therefore one could reasonably defer to using just the local template rather than simulating the scalar-exchange template explicitly. 
An optimal analysis, one that is focused explicitly on a given target template, benefits from simulating that exact template and obtaining predictions for the resulting late-time structure.

Our philosophy on constraining the parameter spaces of these inflationary models, as reflected in our presentation of Figure \ref{fig:Constraints}, is to follow the same approach as analyses of cross-sections in particle physics, particularly in dark matter searches. There, one uses a measurement or experiment to place constraints on an interaction strength given some properties of the DM particle (such as particle mass). In our case, we quote constraining power on $\fNL$ (the ``interaction strength'') given some properties of the template, which are often connected to properties of the particle interaction that generates that template.

While the results of this section focus on a future dataset (LSST Y10), there is significant potential to carry out such lensing-based analyses of PNGs in the near term as well. Existing datasets have similar sky-area coverage and constraining power as the LSST Y1 dataset. A combination of lensing data from the Dark Energy Survey \citep{Gatti2021ShearCatalog, Secco2022Shear, Amon2022Shear} and the Dark Energy Camera All Data Everywhere (DECADE) project \citep{DECADE1, DECADE4, DECADE5} provides 270 million galaxies spanning 13,000 $\deg^2$ of the sky (one-third of the full sky). The redshift distribution for this data \citep{Myles2021PhotoZ, DECADE2} is shallower than LSST Y10, but this is not a limitation as the signal peaks at lower redshifts \citep[][see their Figure 6]{Anbajagane2023Inflation}. Thus, the existing lensing data is already an apt avenue for pathfinder analyses of inflation.

\subsection{Matter Power Spectrum and Bispectrum} \label{sec:results:matter}

Given the constraints above, we now systematically disentangle the origin of the information in the late-time structure. In \citet{Anbajagane2023Inflation}, we showed how the matter and halo fields both respond to the impact of PNGs in connected ways. We perform a similar study here for the range of models discussed in this work. We first explore features in the matter density field; specifically, the power spectrum and bispectrum. The latter is estimated using the approach of \citet{Scoccimarro:2015:Bispectrum}, and the estimator is detailed further in Appendix~\ref{appx:Validation:Bispec}.

\begin{figure}
    \centering
    \includegraphics[width=\columnwidth]{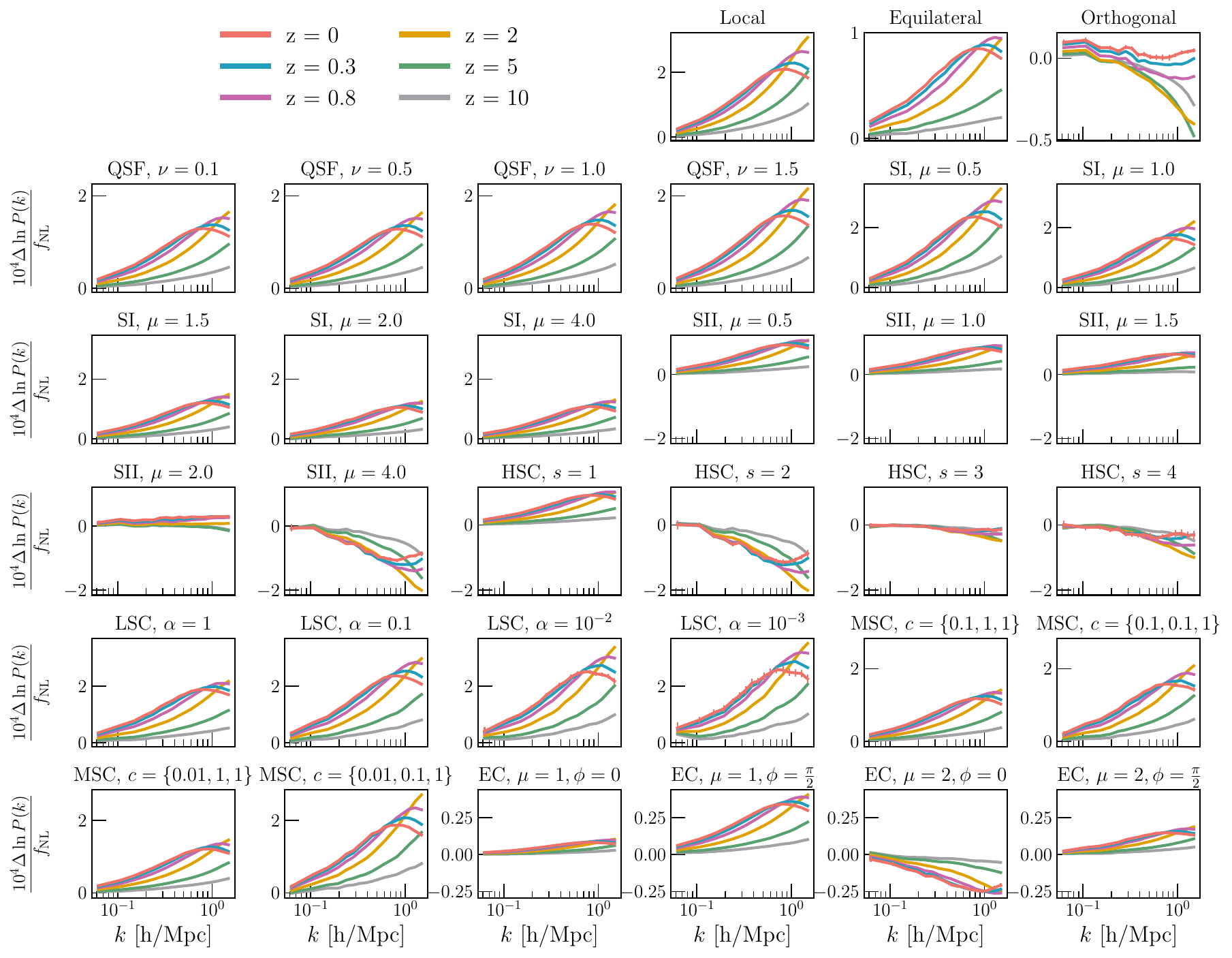}
    \caption{The derivative of the power spectrum with input $\fNL$ value for different templates, and at different redshifts. The range of the y-axis is shared for all variants of a given template (e.g. changes in input mass or spin). For values of $\fNL = 100$, there is a 1\% to 2\% change in the power spectrum across all scales. The impact is generally up to a factor of 2 smaller at higher redshifts ($z = 5$). The error bars show the uncertainty on the mean derivative, obtained from bootstrapping over ten independent realizations. For visibility, we only show uncertainties at $z = 0$. In most panels the uncertainties are smaller than the thickness of the lines.}
    \label{fig:results:mPk}
\end{figure}

Figure \ref{fig:results:mPk} presents the fractional change in the power spectrum for unit $\fNL$. For characteristic values of $\fNL = 100$, we see the signal is around 1\% to 2\%. This is indeed a small change, but with a somewhat characteristic shape across the range of scales probed here. Indeed, it is possible to use just the power spectrum (or Gaussian statistics) to constrain PNGs in weak lensing, but this hinges crucially on tight priors for other cosmology parameters (particularly on $\Omega_m$, the matter energy density fraction at $z = 0$, and $\sigma_8$, the root-mean squared amplitude of fluctuations at $z = 0$ on scales of $8 \mpc/h$). In \citet{Anbajagane2023Inflation}, we show that once $\Omega_m$ and $\sigma_8$ are jointly constrained alongside $\fNL$, the Gaussian statistics provide no significant constraint on $\fNL$. 
Nonetheless, we present the derivative of $P(k)$ to illustrate the changes to the matter density field due to the introduction of PNGs. 
Note that under perturbation theory, the impact of PNGs enter the power spectrum prediction at the one-loop term and scales as $\fNL^2$. 
In this case, we would expect the numerical derivative $P^{\fNL = 100}(k) - P^{\fNL = -100}(k) \approx 0$. The fact that we see a nonzero derivative, to high significance, shows the highly nonlinear (beyond perturbative regime) impact of $\fNL$ on the power spectrum. 

\begin{figure}
    \centering
    \includegraphics[width=1\columnwidth]{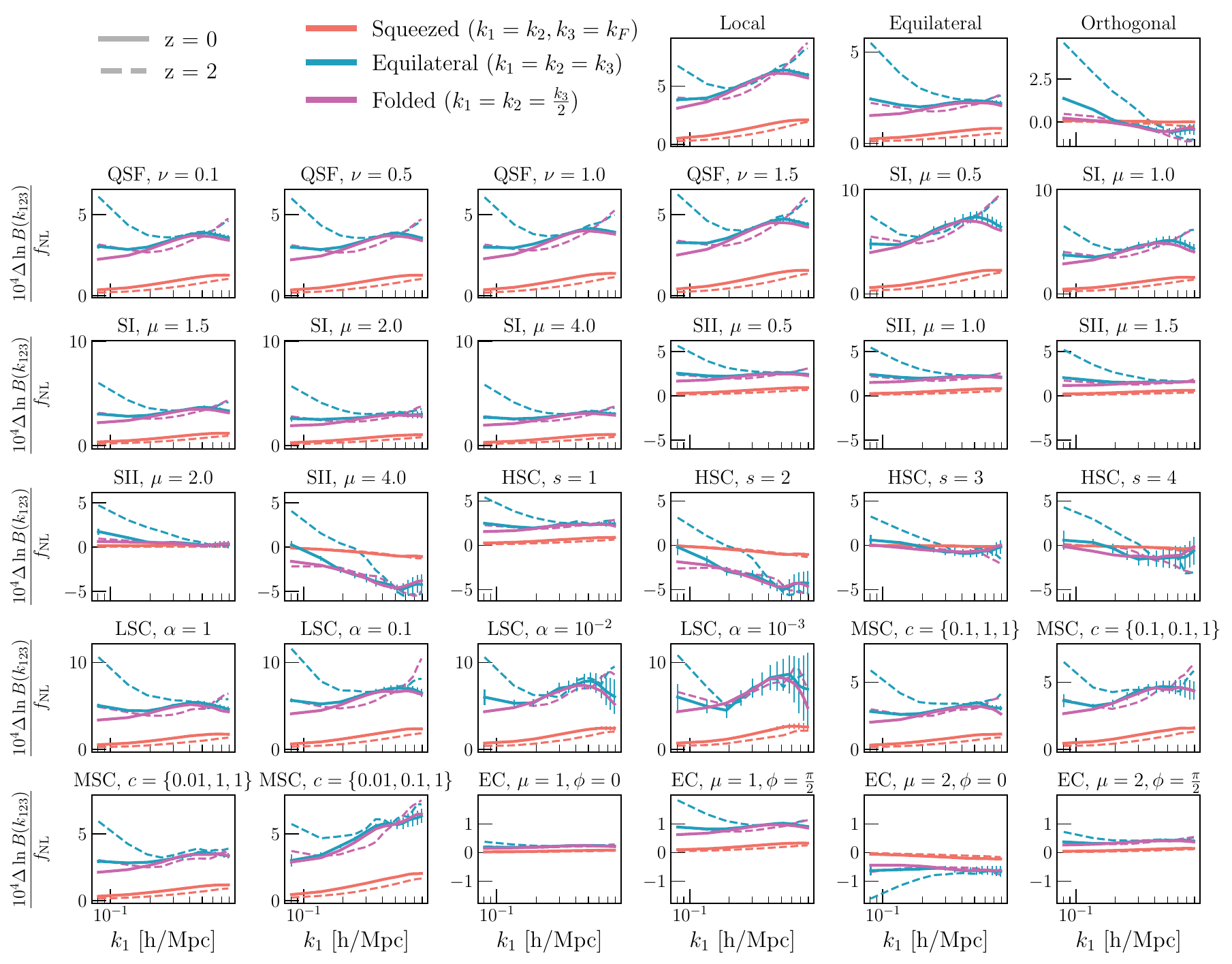}
    \caption{Similar to Figure \ref{fig:results:mPk} but for the matter density bispectrum. There is a 3\% to 5\% change in the bispectrum across all scales for characteristic values of $\fNL = 100$. Given gravitational nonlinearities generate a large bispectrum in the squeezed limit, the relative change due to PNGs is suppressed in that specific limit. The amplitude ($B$ not $\Delta B$) of the late-time bispectrum peaks in the squeezed limit. For visibility reasons, we only show uncertainties for the equilateral and squeezed limits at $z = 0$. The measurements are smoothed with a narrow Gaussian kernel for visualization.}
    \label{fig:results:mBk}
\end{figure}

Figure \ref{fig:results:mBk} then presents the fractional change in the bispectrum for unit $\fNL$. Here, we see somewhat larger changes relative to that seen in the power spectrum measurement. In particular, we see the spectra change by 3--5\% for characteristic values of $\fNL = 100$. The fractional change is largest in the equilateral and folded configurations, i.e., it is suppressed in the squeezed limit relative to the other types. This is because the matter bispectrum has a larger amplitude (by more than an order of magnitude) in the squeezed limit due to the impact of nonlinear structure \citep[][see their Figure 5]{Takahashi:2020:BiHaloFit}, so the fractional change due to PNGs is smaller as a result.

The results in Figure~\ref{fig:results:mBk} also show that the late-time matter bispectrum has fairly regular shapes across the different models. The largest scales used in our measurements are already quasi-linear ($k > 0.1 \,\,h/\mpc$) and on these scales, the mixed term between the primordial $\fNL$ signal and quasi-linear perturbation theory signal dominates over the primordial signal and damps most of the unique signature from the primordial bispectra. A version of this is shown in \citet{Goldstein:2025:CosmoColl} for oscillatory bispectra in the squeezed limit. Hence, most of our measured bispectra follow a fairly similar (but not the same) scale dependence.

\subsection{Halo Abundance and Bias} \label{sec:results:halos}

We can then check the impact of PNGs on statistics of massive halos. Given the resolution of our simulation, a halo of $\Mtwohc = 10^{13.5} \msun/h$ is resolved by roughly 30 particles. For the purposes of measuring halo counts and halo clustering, this resolution is adequate. Any studies on the internal structure of these halos will need to be limited to much higher masses. Studies of halo statistics are particularly relevant as \citet{Anbajagane2023Inflation} identified this as the primary origin of the information in the weak lensing field, supporting prior work indicating the information content in the halo statistics \citep{Dalal2008ScaleDependentBias, Shirasaki2012fNL, Marian2011fNL, Hilbert2012fNL, Jung2023fNLHMFQuijote}.

\begin{figure}
    \centering
    \includegraphics[width=\columnwidth]{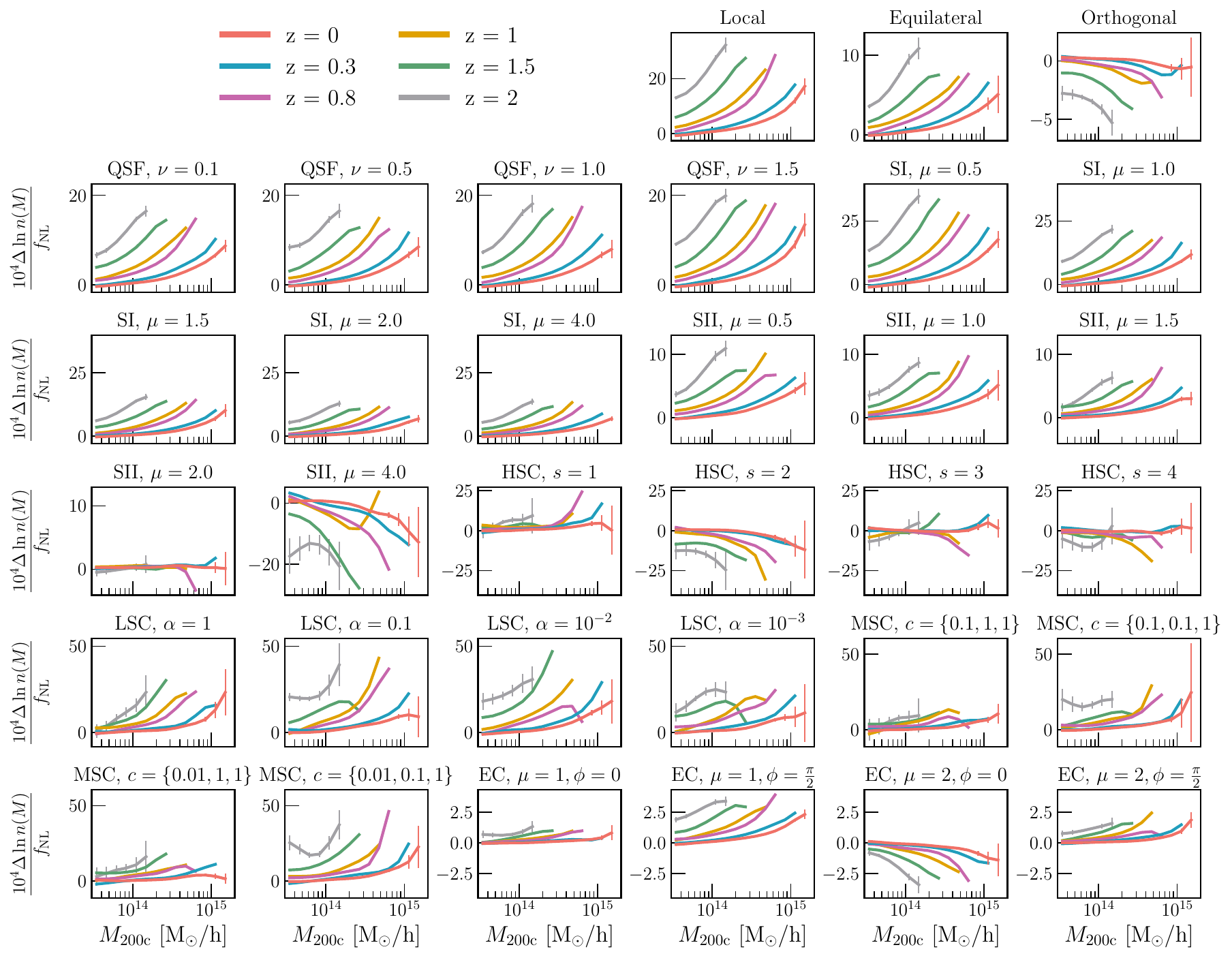}
    \caption{Similar to Figure \ref{fig:results:mPk} but for the halo mass function (HMF). A value of $\fNL = 100$ results in an order 10\% to 25\% change in the halo mass function in many models; an order of magnitude larger than the 2-3\% change seen in the power spectrum and bispectrum. The effect grows with redshift (at fixed mass) because the peak height (or ``rarity'') of a given halo mass increases with redshift. For visibility we only show the uncertainties at $z = 0$ and $z = 2$. The measurements are smoothed with a narrow Gaussian kernel for visualization.}
    \label{fig:results:HMF}
\end{figure}

Figure \ref{fig:results:HMF} presents the change in the halo mass function for a unit change in $\fNL$. As shown in \citet{Anbajagane2023Inflation}, the abundance of halos changes with $\fNL$. The amplitude of the change is the strongest of all effects considered thus far, as it is between 10--25\% for characteristic values of $\fNL = 100$. The origin of this signal is because PNGs alter the shape of the initial density distribution, where this alteration has a significantly stronger impact on the tails of this distribution. The upper tail encompass regions of space that eventually collapse to form halos, hence the coupling of $\fNL$ with halo abundance.\footnote{In principle, the statistics of voids---which form from the negative (not positive) tail of the density field---should also be affected by $\fNL$, and this change will be traced by the weak lensing observations. We have not generated void catalogs in our simulations and are therefore unable to answer this question.} The exact sign of the change depends on the normalization of the template (as the normalization can induce a negative sign in the final template), but in general positive $\fNL$ results in an enhanced growth of massive halos. At fixed mass, this enhancement grows with redshift as the peak height (or rarity) of a halo increases with redshift, and PNGs have a larger impact on the abundance of rarer objects. These objects occur as ``peaks'' in the weak lensing convergence field and can be probed via a range of higher-order statistics, including via the combination of the second and third moments used in this work.

\begin{figure}
    \centering
    \includegraphics[width=\columnwidth]{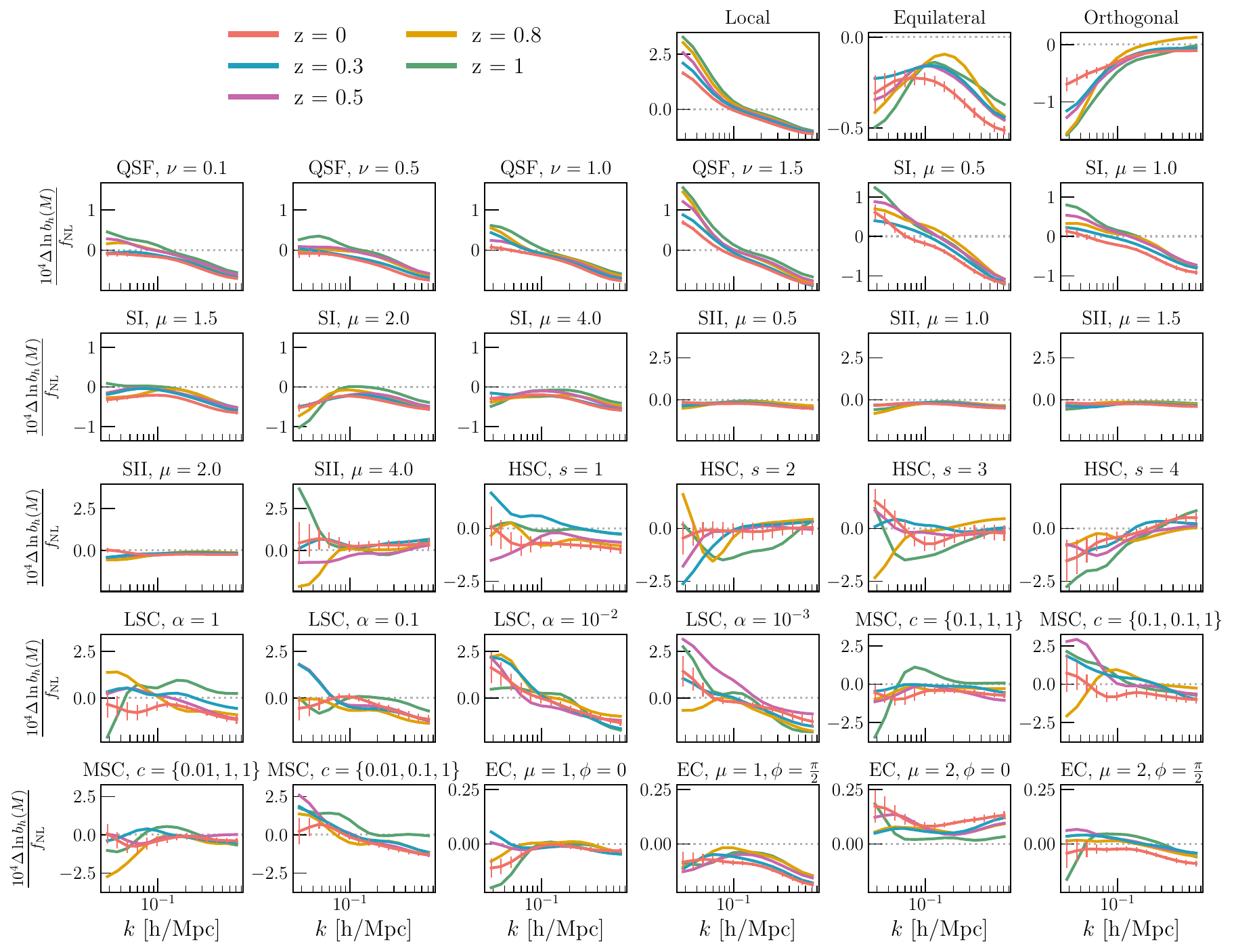}
    \caption{Similar to Figure \ref{fig:results:mPk} but for the halo bias, estimated as $P_{\rm hm}/P_{\rm mm}$. The Local template produces its characteristic, growing bias on large scales, while other templates show much milder scale-dependence in their bias. We estimate this for a sample with $M_{\rm 200c} > 10^{14} M_\odot/h$. We only show the error bars at $z = 0$ for enhanced visibility. The measurements are also smoothed with a narrow Gaussian kernel for visualization. The somewhat rapid fluctuations of some of the high-redshift curves are due to inherent noise in the measurement. We have confirmed the minimal impact of shot noise in these estimates by confirming their statistical consistency with measurements derived from a random 50\% subsample of halos.}
    \label{fig:results:halobias}
\end{figure}

Figure \ref{fig:results:halobias} presents the change in halo bias for unit change in $\fNL$. We estimate this by measuring the halo-matter and matter-matter power spectrum, $P_{\rm hm}$ and $P_{\rm mm}$, respectively. Then we compute the bias as $b_h = P_{\rm hm} / P_{\rm mm}$. The change in bias is computed as $\Delta \ln b_h = (b_h^{ +\fNL} - b_h^{ -\fNL}) / (2 b_h^{ \fNL = 0})$. The first row reproduces the seminal result of \citet{Dalal2008ScaleDependentBias}---the halo bias in local-type $\fNL$ divergences towards large scales. In the equilateral type, the bias is scale-independent on larger scales (while there is an apparent trend in the lines, it is consistent with scale-independent bias within the computed error bars) and shows mild scale-dependence towards the smaller scales. 

The scalar-exchange templates (QSF, SI, SII) broadly follow the behavior of the local, equilateral, orthogonal types, though there are still minor differences across the templates. The QSF templates show similar behaviors since they are quite correlated with the local template. The SI template mimics the local type for low mass, $\mu < 1$, and the equilateral type for high mass, $\mu > 1$. This follows the simple expectations from visually analyzing the similarities of the templates in Figure \ref{fig:Template}, where the SI template is similar to the local and equilateral templates at low and high $\mu$, respectively. A similar behavior is seen in SII, where it mimics the equilateral template at low mass and the orthogonal template at high mass. In general, we find all templates show some kind of scale-dependence in their bias, and there is also significant scale-dependence on small-scales (larger $k$). The templates with nonzero spin and sound-speed deviations show noisier estimates, particularly at higher redshifts.

\section{Conclusions}\label{sec:conclusions}

Primordial Non-Gaussianities (PNGs) are signals imprinted in the initial conditions of our Universe that encode the dynamics of the inflationary epoch. 
At leading order, these signals are captured by the bispectrum in Fourier space, whose precise form depends on the nature of the inflaton sector---arising, for example, from its self-interactions or couplings to additional fields. 
Over the past two decades, a handful of simple PNG models have been explored through simulations, but the vast majority of signatures---including those associated with cosmological collider physics---have not been probed.

In this paper, we presented a method for generating initial conditions (ICs) with arbitrary bispectra, including those from collider models. 
We then generated the first suite of simulations that propagate the impact of thirty different primordial bispectra into the deeply nonlinear regime of structure formation. We provided extensive validation of our IC generation method (Appendix \ref{appx:Validation}) and showcase its robustness from numerical artifacts.

\noindent Our main results are summarized as follows:

\begin{itemize}
    \item We can generate initial conditions with arbitrary bispectra (corresponding to different PNG models) and resolve their predictions for nonlinear structure. The transformations performed on the (Gaussian) initial conditions to achieve this are all well-behaved and do not induce numerical artifacts (Appendix \ref{appx:Validation}).
    
    \item As found in previous works, the primary signature of PNGs in the nonlinear density field is through changes to the abundance of massive halos (Figure \ref{fig:results:HMF}) and to the clustering of these halos (Figure \ref{fig:results:halobias}). This naturally results in differences in the nonlinear matter power spectrum (Figure \ref{fig:results:mPk}). However, the impact on the halo abundance is the strongest out of all of these effects.
    
    \item The matter bispectrum also shows changes with different input $\fNL$, with the largest fractional change occurring in the equilateral and folded limits (Figure \ref{fig:results:mBk}).
    
    \item We forward-model the observed (noisy, masked) lensing field from the LSST Y10 dataset, and find that the lensing-only constraints---using the second and third moments of the convergence field as a summary statistics---are at most within 30\% to 50\% of the current leading constraints from the \textit{Planck} 2018 temperature and polarization bispectra (Figure \ref{fig:Constraints}). This showcases that lensing is a worthwhile addition, with highly complementary information, to the list of late Universe probes used to constrain PNGs. 

    \item The late-time signatures of the different templates do exhibit some correlations, but many are only weakly correlated with other templates, including being weakly correlated with the standard local, equilateral, and orthogonal types (Figure \ref{fig:Corr}).
\end{itemize}

Regarding future improvements, there is significant potential in increasing the information extracted from the lensing convergence field. In particular, we have used a somewhat simplistic set of statistics in this work. Incorporating others, such as those used in the analyses of \citet{Gatti:2024:WPH, Prat:2025:Homology}, can significantly improve on our constraining power. In particular, persistent homology statistics used in \citet{Prat:2025:Homology} more cleanly encapsulate the signatures in peaks and voids of the convergence field, and so can be a more natural statistic for the signatures from $\fNL$ \citep{Cole:2020:HomologyPNG}. 

Turning to the formalism introduced in our work, our template approximation was built on a flexible choice of basis functions.
For optimal results, the basis functions should satisfy several properties: (i) separability in $\kOne, \kTwo, \kThree$, (ii) sufficient orthogonality in the three-dimensional construction, $Q =  g_i(\kOne) g_j(\kTwo) g_k(\kThree)$, (iii) functional form that is a natural fit for the templates being considered, and (iv) no divergent behaviors in the one-loop primordial power spectrum. 
Of these, property (ii) is the most subtle, since reproducing smooth template behavior can sometimes require fine-tuned cancellations among basis terms. 
The present method works well for our purposes, and could be generalized further through more ideal choices of basis functions.

Analyses of $\Lambda$CDM cosmology have benefited significantly from the use of multiple cosmological probes, each probe with its own strengths and shortfalls. A similar approach has not been possible for PNG-extended cosmologies, primarily because of the lack of reliable simulations to calibrate probes that rely on the nonlinear nature of structure formation (\eg weak lensing, cluster abundances, etc.). This work builds on previous efforts and presents a generalized technique for producing simulations with any primordial bispectra. As a result, studies of PNGs---and of cosmological collider physics in particular---can now pursue multi-probe analyses. 
This will enable more precise constraints from the late Universe, allow for more meaningful comparisons with constraints from the early Universe, and help discover useful synergies between probes that can be exploited in our pursuit of PNGs. 
Simulations have been a critical tool underpinning much of the current landscape of precision cosmology, and the analyses of PNGs will only flourish as this tool is used to tackle constraints on primordial physics.

\section*{Acknowledgements}

DA thanks many people---Austin Joyce, Josh Frieman, Lucas Secco, Olivier Dore, Ben Wandelt, Lam Hui, Sam Goldstein, Oliver Philcox, Chihway Chang, Bhuvnesh Jain, Wayne Hu, Katrinn Heitmann, Salman Habib, Will Coulton, Paco Villaescusa-Navarro, Chun-Hao To, Nick Kokron, Moritz Munchmeyer, and Mat Madhavacheril---for enlightening conversations across the two years leading up to the publication of this work. DA is supported by the National Science Foundation Graduate Research Fellowship under Grant No. DGE 1746045. HL is supported in part by the DOE award DE-SC0013528.

All analysis in this work was enabled greatly by the following software: \textsc{Pandas} \citep{Mckinney2011pandas}, \textsc{NumPy} \citep{vanderWalt2011Numpy}, \textsc{SciPy} \citep{Virtanen2020Scipy}, and \textsc{Matplotlib} \citep{Hunter2007Matplotlib}. We have also used
the Astrophysics Data Service (\href{https://ui.adsabs.harvard.edu/}{ADS}) and \href{https://arxiv.org/}{\texttt{arXiv}} preprint repository extensively during this project and the writing of the paper.

\section*{Data Availability}

Our pipeline for generating initial conditions---including our technique for approximating bispectra using separable terms---is available at \url{https://github.com/DhayaaAnbajagane/Aarambam}. The simulations are publicly released as part of the \textsc{Ulagam} suite. More details can be found at \url{https://ulagam-simulations.readthedocs.io}. The released products include the 2D density shells, 3D density fields, and halo catalogs for all snapshots. Please contact DA if you are interested in other products from the simulations.

\bibliographystyle{mnras}
\bibliography{References}


\newpage
\appendix

\section{Generating Initial Conditions} \label{appx:ICs}

In Section \ref{sec:sims:ICs}, we extensively discussed our method for generating non-Gaussian initial conditions for arbitrary non-separable bispectra. In this appendix, we provide additional technical details on this method, including the handling of IR-divergent terms in Appendix \ref{appx:ICs:Divergent} and of numerical accuracy in Appendix \ref{appx:ICs:TechDetails}.

\subsection{Cancellation of Divergences} \label{appx:ICs:Divergent}

In Section \ref{sec:sims:ICs}, we introduced our method for generating initial conditions (ICs) corresponding to arbitrary bispectrum templates. A requirement in this method is to ensure that the choice of kernel, $K$, does not lead to divergent terms in the power spectrum (relative to the tree-level power-spectrum prediction). To cancel such divergent terms, we can use linear combinations of symmetric constructions of a given kernel, and choose the amplitudes in this combination such that the divergences are cancelled. This follows the approach taken in \citet{Scoccimarro2012PNGs} for removing divergences in the kernels of the local, equilateral, and orthogonal templates.

We start by considering three classes of templates, distinguished by the allowed symmetries of the indices. The first is a template $K^{abc}$ with the indices $a = b = c$, which results in no additions since the symmetry of the template enforces only a single, unique template. The second is for $a = b \neq c$, where the template can be written as the linear combination of two terms
\begin{align}
    K^{a = b\neq c}_{12} = & \,\,(1 - u_{abc})\kThree^{c - 3}\kOne^{b}\kTwo^{b}  + \frac{1}{2}u_{abc}\kThree^{b - 3}(\kOne^{c}\kTwo^{b} + \kOne^{b}\kTwo^{c})\,,\nonumber\\
    K^{a \neq b =  c}_{12} = & \,\,\frac{1}{2}(1 - u_{abc})\kThree^{b - 3}(\kOne^{a}\kTwo^{b} + \kOne^{b}\kTwo^{a})+ u_{abc}\kThree^{a - 3}\kOne^{b}\kTwo^{b}\, .
\end{align}
We have explicitly written out the two versions of the linear combinations, differing in the term corresponding to $u_{abc}$ and that corresponding to $1 - u_{abc}$, which in turn depends on which of the two indices are symmetric/repeated in the kernel. Via this choice, the amplitude $1 - u_{abc}$ always corresponds to the template that is part of our basis functions, $Q$, and the amplitude $u_{abc}$ to the additional, permuted kernel whose inclusion is being considered. Finally, the third class is for templates where $a \neq b \neq c$,
\begin{align}
    K^{a \neq b\neq c}_{12} = & \,\,\frac{1}{2}(1 - t_{abc} - s_{abc})\kThree^{c - 3}(\kTwo^{b}\kThree^{a} + \kTwo^{a}\kThree^{b}) \nonumber\\ 
    & + \frac{1}{2}t_{abc}\kThree^{b - 3}(\kTwo^{a}\kThree^{c} + \kTwo^{c}\kThree^{a}) + \frac{1}{2}s_{abc}\kThree^{a - 3}(\kTwo^{c}\kThree^{b} + \kTwo^{b}\kThree^{c})\,.
\end{align}
The factors of $1/2$ in certain terms above are just a definitional choice. Any symmetry factors are included in $\alpha_{ijk}$ and must not be double-counted elsewhere. For all the above, we continue with our constraint $a \leq b \leq c$. The coefficients, $u_{abc}, t_{abc}, s_{abc}$ can be constrained by computing $K^2_{12}$ and inspecting terms with divergences in $\kThree$; see Eq.~\eqref{eqn:ICs:1loop} and also Appendix~D of \citetalias{Scoccimarro2012PNGs}.

\noindent We focus on divergences that scale like, or more steeply than, the tree-level power spectrum, which limits us to $P^{\text{1-loop}} \propto k^x$ with $x \leq -3$. To cancel the $k^{-6}$ and $k^{-5}$ terms in the one-loop power spectrum, we must enforce the following constraint:
\begin{align}
    0= & \,\,\kTwo^{6} {\alpha}_{033} {u}_{033} + \kTwo^{5} {\alpha}_{023} {s}_{023} + \kTwo^{4} {\alpha}_{013} {s}_{013} + \kTwo^{4} {\alpha}_{022} {u}_{022} + \kTwo^{3} {\alpha}_{003} {u}_{003} + \kTwo^{3} {\alpha}_{012} {s}_{012} \nonumber\\
    & + \kTwo^{2} {\alpha}_{002} {u}_{002} + \kTwo^{2} {\alpha}_{011} {u}_{011} + \kTwo {\alpha}_{001} {u}_{001} + {\alpha}_{000}\,.
\end{align}
This leads to a fairly simple constraint on the parameters
\begin{equation}
    \alpha_{000},\, u_{001},\, s_{023},\, u_{033}  = 0\,,\nonumber
\end{equation}
and
\begin{equation}
    {u}_{022}  = -\frac{\alpha_{013} {s}_{013}}{\alpha_{022}} = 0\,,\quad 
    {s}_{012} = -\frac{\alpha_{003} {u}_{003}} {\alpha_{012}} = 0\,,\quad
     {u}_{011}  = -\frac{\alpha_{002} {u}_{002}}{\alpha_{011}} = 1\,.
\end{equation}
We have further simplified the values of $u_{022}$ and $s_{012}$ given we can take $s_{013}, u_{003} = 0$ without any consequences to the divergent behavior. We set ${u}_{011} = 1$ after finding the constraint for ${u}_{002} = -\frac{\alpha_{011}}{\alpha_{002}}$ further below. The statement $\alpha_{000} = 0$ is not a constraint on a free parameter and is instead an indication that we must remove the function, $(\kOne\kTwo\kThree)^{-3}$, from our basis set as otherwise the divergences to the one loop power spectra are unavoidable. In practice removing this term has no effect in this work as the templates we study do not have features with such a scaling in all three wavevectors. We have also explicitly checked that the bispectrum templates used in this work are adequately fit even without including this term. The constraint $s_{012} = -\alpha_{003} {u}_{003} / \alpha_{012}$ corresponds to the result, $s = -u/2$, found in \citetalias{Scoccimarro2012PNGs} for the equilateral and orthogonal templates.

We can now do the same exercise for the $k^{-4}$ and $k^{-3}$ terms, which leads to a second constraint equation
\begin{align}
    0&=  \kTwo^{6} {\alpha}_{133} {u}_{133} + \kTwo^{5} {\alpha}_{123} {s}_{123} + \kTwo^{4} {\alpha}_{113} {u}_{113} + \kTwo^{4} {\alpha}_{122} {u}_{122} + \kTwo^{3} {\alpha}_{013} {t}_{013} \nonumber\\
    &\quad +  \kTwo^{3} {\alpha}_{112} {u}_{112} + \kTwo^{2} {\alpha}_{012} {t}_{012} + \kTwo^{2} {\alpha}_{111} + \kTwo {\alpha}_{002} {u}_{002} + \kTwo {\alpha}_{011} + {\alpha}_{001}\,,
\end{align}
which is now translated into the following constraints on the parameters,
\begin{align}
    \alpha_{001}, s_{123}, u_{133}  = 0\,,\quad
    u_{122}  = -\frac{\alpha_{113}u_{113}}{\alpha_{122}} = 0\,,\nonumber\\
    u_{002}  = -\frac{\alpha_{011}}{\alpha_{002}}\,,\quad
    t_{012}  = -\frac{\alpha_{111}}{\alpha_{012}}\,,\quad
    t_{013}  = -\frac{\alpha_{112}u_{112}}{\alpha_{013}} = 0\,.
\end{align} 
Analogous to the previous case, we see a requirement of $\alpha_{001} = 0$, which is enforced by dropping the term $\kOne^2\kTwo^2\kThree^1$. We have once again simplified the expressions for certain parameters---namely $u_{122}$ and $t_{013}$---where allowed. Note that the constraint $t_{012} = -\frac{\alpha_{111}}{\alpha_{012}}$ matches the constraint from \citetalias{Scoccimarro2012PNGs} which was used to null the $k^{-4}$ divergences in the equilateral and orthogonal templates. 

All other $u_{abc}, t_{abc}, s_{abc}$ parameters not included in the constraint equations above can be set to $0$ without any consequences for the divergent behaviors of the template. We reiterate that the modified Legendre polynomials used in our work (see Eq.~\ref{eqn:modefunc}) do not contribute divergences stronger than $k^{-2}$ in the one-loop term. Without our modification \textit{all} Legendre terms contribute at least $k^{-4}$ divergences to the one-loop term.

\subsection{Numerical Considerations}\label{appx:ICs:TechDetails}

There are also a number of subtleties related to the matrix inversion process that we detail here for clarity. These are primarily caused by the highly correlated nature of the three-dimensional basis set; note that the one-dimensional functions are still fairly orthogonal to each other, but the compositions of these functions into a three-dimensional basis significantly amplifies the correlations. In particular, we empirically find that round-off errors---which are generated by this highly correlated basis set---can accumulate and result in divergent behaviors in the one-loop power spectrum of the ICs. We resolve this through a variety of methods:

\textbf{First}, we do not rely on the default \texttt{double} precision (64 bits) of the \texttt{FFTW3} routines \citep{Frigo:2005:FFTW3} used to calculate FFTs. Instead, we use \texttt{long double} precision (128 bits) to separately accumulate the real and imaginary pieces of the gaussian, inflaton potential field. This is a simple way of doubling the precision available for the calculations and alleviates some round-off errors. In practice, this change is not a critical necessity but still a helpful one.

\textbf{Second}, and more importantly, we perform a guided subselection of the modes. This subselection reduces the number of modes in the final basis set that are highly correlated with each other. We tested a simple, random subsampling---akin to the ``dropout'' technique used in training neural networks \citep{Hinton:2012:Dropout}---and found that it works well but is not easily generalizable; to retain good accuracy in the approximation of a given bispectrum, the number of dropped modes is somewhat dependent on the template being decomposed. Furthermore, the correlation between modes increases for larger indices $i,j,k$, which is information not used in a random dropout procedure. Thus, we instead use an iterative approach following the Orthogonal Matching Pursuit (OMP) method \citep{Pati:1993:OMP, Tropp:2007:OMP}. 

In this method, we use a set of modes $\{M_i\}$ to decompose the shape, $S$, and compute the residual $R = S - \sum_{m} \alpha_m \times m$, with the modes $m  \in \{M_i\}$ and coefficients $\alpha_m$. This residual, $R$, quantifies the error in the approximation to the shape, integrated over the full domain. We then find a mode \textit{not in} $\{M_i\}$ that has the strongest correlation with $R$. This is quantified using the absolute value of the inner product in Eq.~\eqref{eqn:ICs:innerproduct}. We then extend the basis to $M_{i + j} = \{M_i\} + \{m_j\}$, using the newly selected mode, $m_j$. We add modes to the basis iteratively until the condition $\langle R, R \rangle / \langle S, S \rangle < \epsilon_R$ is met. We find that $\epsilon_R = 5\times10^{-4}$ provides adequate results. We also add an additional condition that we will use at most $N_{\rm modes}^{\rm max}$ modes from the basis. In practice, we use $N_{\rm modes}^{\rm max} \leq 150$ with the number being lowered for simple templates such as the QSF model (Eq.~\ref{eqn:template:qsf}). Note that the coefficients $\alpha_m$ are recomputed everytime a new mode is added to the basis set.

\textbf{Third,} we further supplement the OMP method to avoid dramatic increases to the conditioning number of the inner product matrix, $\langle Q_i, Q_j \rangle$, used in Eq.~\eqref{eqn:ICs:matrix_eqn}. When using all possible modes, this matrix is ill-conditioned. This leads to significant numerical errors in the standard execution of our pipelines, and sometimes amplifies to catastrophic errors (\eg adding a few extra modes leads to completely nonsensical approximations to the target bispectrum). Therefore, for a given candidate basis function, $m_j$, we compute the matrix condition number before and after the function is added to set $\{M_i\}$. We use the fractional difference in condition numbers to remove any candidate mode, $m_j$, that enhances the risk of an ill-conditioned basis. This criteria is the most critical in preventing catastrophic errors in our basis approximation. We empirically find there are multiple subsets of the basis functions that provide equally adequate approximations to the target bispectrum, but some have large condition numbers and therefore result in visible numerical errors in the resulting initial conditions.

In practice, we first select the top ten functions based on their correlations with the residual, $R$. Of those ten functions, we select the one that minimizes the increase to the condition number. We find empirically that under this setup, the approximation continues to improve as we add more modes but the increase in the condition number is still kept under control.

The above discussion highlights the limitations in finding basis functions that match our requirements. In the ideal case, the basis should be close to orthonormal in the three-dimensional volume, while also being factorizable in $\kOne, \kTwo, \kThree$. The latter requirement often leads to the final three-dimensional basis being highly correlated, as basis functions that are orthonormal in 1D are correlated when multiplied together to form a 3D function. In practice, one relies on numerical methods to orthogonalize the 3D basis functions. This will then lead to the same numerical issues we have identified in this work. As repeated before in our discussions, improving choice of basis functions has the potential to greatly generalize our methods to a wider variety of models.

\section{Validation of Initial Conditions}\label{appx:Validation}

The transformations we perform on the Gaussian initial conditions can potentially lead to numerical artifacts that then bias the resolved structure in the $N$-body simulation. Therefore, we perform a number of key validations to ensure we can robustly use our method of generating ICs. We perform these validations on a grid of $128^3$ in a box with $L = 1 \gpc/h$. The grid spacing is sparse (compared to our fiducial setup) due to computational cost --- all validations below require a large number of three-dimensional FFTs. So the computational complexity goes as $\propto N_{\rm grid}^3 \log(N_{\rm grid})$. We have verified that using a grid of $256^3$ does not change our conclusions. All tests below show the average over 30 independent realizations. We also generate pairs of fields with the same random phases but with $\fNL = \pm X$ and use them to suppress cosmic variance in our final estimates.

\subsection{Bispectrum}\label{appx:Validation:Bispec}

One of the key validations of this work is measuring the bispectrum of the modified inflaton potential field and confirming that it matches the theoretical expectation. Given the formal definition of a bispectrum, as $B(\kOne, \kTwo, \kThree) = \langle X(\kOne)X(\kTwo)X(\kThree)\rangle$ for field $X$, the naive estimator of this quantity scales as $N_{\rm grid}^9$. Instead, we use the estimator of \citet{Scoccimarro:2015:Bispectrum}, which we briefly detail below.

\noindent The \textit{binned} bispectrum estimator can be written as,
\begin{equation}\label{eqn:BispecEst}
    B(\kOne, \kTwo, \kThree) = \frac{1}{A}\int_{\Delta \kOne} d\kOne^3 \int_{\Delta \kTwo} d\kTwo^3\int_{\Delta \kThree} d\kThree^3\,\,X(\kOne)X(\kTwo)X(\kThree) \,\delta(\vec k_1 + \vec k_2 + \vec k_3)\,,
\end{equation}
where $\Delta k_i$ are the bins for wavevector $i \in \{1, 2, 3\}$. We can rewrite the delta function as 
\begin{equation}
    (2\pi)^3\delta(\vec k_1 + \vec k_2 + \vec k_3) = \int d^3x\, e^{-i\vec{x}\cdot(\vec k_1 + \vec k_2 + \vec k_3)}\,.
\end{equation}
Through this substitution, the 9-dimensional integral above (which leads to the $N_{\rm grid}^9$ scaling) can be done through separable Fourier transforms,
\begin{equation}\label{eqn:BispecEstFFT}
    B(\kOne, \kTwo, \kThree) = \frac{1}{A} \int d^3x\,\,\, {\rm FFT}[W_{\Delta \kOne}X(\kOne)] \times {\rm FFT}[W_{\Delta \kTwo}X(\kTwo)] \times {\rm FFT}[W_{\Delta \kThree}X(\kThree)]\,,
\end{equation}
where $W_{\Delta \kOne}$ is a Fourier-space binning kernel---it is unity in the range $k_{i} < \kOne < k_{i + 1}$, where $k_i$ are the bin-edges, and zero everywhere else. For computational reasons, we compute the bispectra in just the squeezed, equilateral, and folded limits. The normalization $A$ is computed by removing the field, $X$, in Eq.~\eqref{eqn:BispecEstFFT} and performing the FFTs over just the window functions. This evalutes the number of triangles per choice of $\kOne, \kTwo, \kThree$.

For a consistent comparison between theory and measurement, the theoretical bispectrum model must also be binned in a consistent manner. This is done as
\begin{equation}
    B_{\rm th, bin}(\kOne, \kTwo, \kThree) = \frac{1}{A}\int_{\Delta \kOne} d^3k_1 \int_{\Delta \kTwo} d^3k_2\int_{\Delta \kThree} d^3k_3\,\,B_{\rm th}(\kOne, \kTwo, \kThree) \,\,\delta(\vec k_1 + \vec k_2 + \vec k_3)\,,
\end{equation}
which is once again a computationally expensive integral, but can be done using FFTs \textit{as long as $B_{\rm th}$ is separable.} Given this separability is a requirement for generating initial conditions with the given bispectrum template, we can simply use the results from the method detailed in Section \ref{sec:sims:ICs} to obtain $B_{\rm th}$ as a sum of separable terms. Then, $B_{\rm th, bin}$ can be estimated using the same technique as is used to measure the data.

\begin{figure}
    \centering
    \includegraphics[width=\columnwidth]{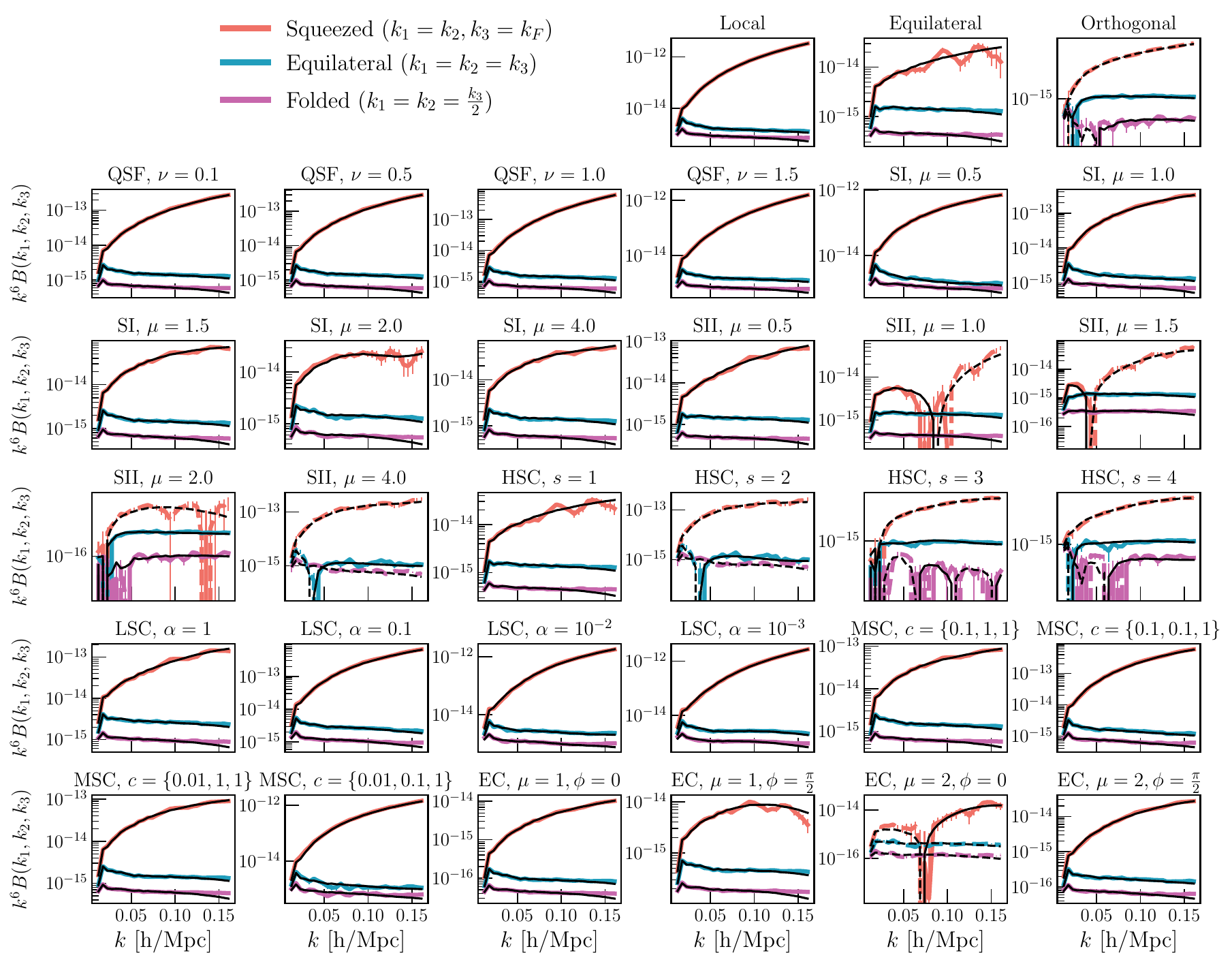}
    \caption{The average bispectrum (over 30 realizations) of the ICs compared to the theoretical expectation. We compute the measurement as the difference, $B(k) = (B^{\fNL = +X}(k) - B^{\fNL = -X}(k))/2$, which is noise-suppressed. The errorbars show the uncertainty on the mean bispectrum and are estimated using bootstrap realizations. In all cases, the measured bispectrum is in agreement with the theoretical expectation. We note a slight suppression in power in the folded limit at large wavenumbers. We have confirmed this effect is due to resolution and vanishes if we double the resolution of the grid used for the calculations.}
    \label{fig:Validation:Bispec}
\end{figure}

Figure \ref{fig:Validation:Bispec} presents the bispectrum of the ICs, averaged over 30 realizations, and compares it to the binned, theoretical expectation. In all cases we find good agreement between the two. Note that we estimate the Bispectrum in the data as $B(k) = (B^{\fNL = +X}(k) - B^{\fNL = -X}(k))/2$, which suppresses the noise in the estimate. The errorbars in Figure \ref{fig:Validation:Bispec} are obtained by a simple bootstrap procedure, and are errors on the mean. There is a tendency for the theory estimates to be slightly lower than the data at high $k$, particularly for the folded limit, and we have verified this is an effect of the grid resolution. All estimates above are made on grids of $128^3$, and we find this suppression is not observed if we increase the resolution to $256^3$. However, the latter is computationally expensive so we do not redo this calculation at this higher resolution. The existing approach is sufficiently accurate to show our ICs exhibit the correct bispectra. While Figure \ref{fig:Template} showed non-trivial, oscillatory features in some bispectrum templates, these are not easily found in the measured bispectra, partially because of the impact of binning but also because the range in Figure \ref{fig:Template} is up to $k = 1 \, {\rm h/Mpc}$ whereas in the above plot we are limited to $k \approx 0.16 \, {\rm h/Mpc}$ due to the resolution of the grid.

\subsection{Power Spectrum}\label{appx:Validation:Pk}

Once the correct bispectrum is generated in the initial conditions, one must be careful in ensuring this modification to the initial potential field does not result in any divergent corrections to the power spectrum of the field. In order to preserve the perturbative nature of the model, the addition of primordial non-Gaussianities should not impact the tree-level power spectrum.

In this section, we confirm that our different bispectrum do not have any divergent one-loop corrections. In practice, we would like any corrections to the power spectrum to be below 1\%. This limit is smaller than the current $1\sigma$ uncertainty on the amplitude of the primordial power spectrum \citep{Planck:2020:CosmoParams}. We stress that in practice, most of our chosen models are orders of magnitude below this requirement.

\begin{figure}
    \centering
    \includegraphics[width=\columnwidth]{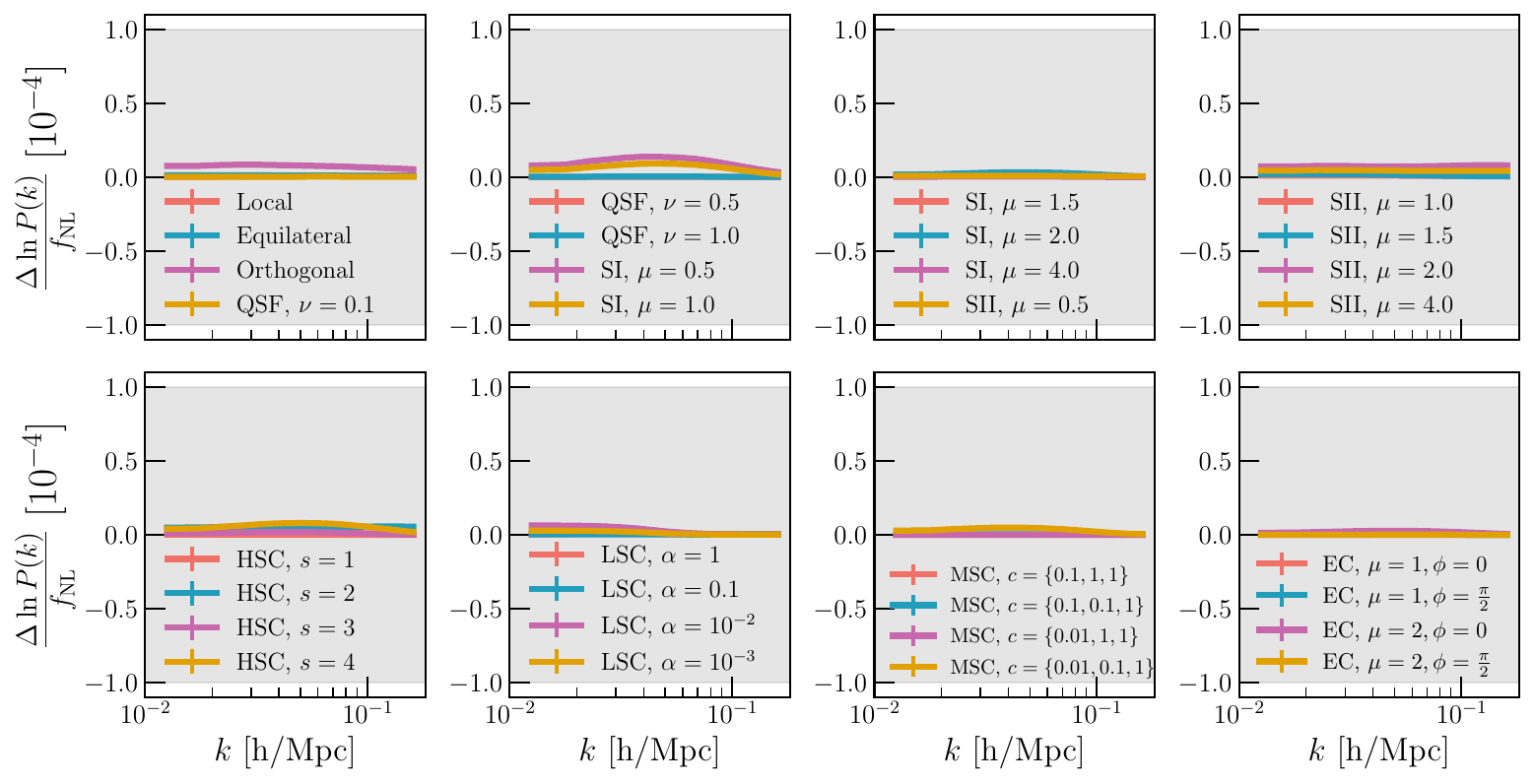}
    \caption{Measurements of the one-loop corrections (averaged over 30 realizations) to the power spectrum due to adding a given bispectrum model to the initial conditions. If the model has a relative change of $\Delta\ln P < 10^{-4}$ then a field with $\fNL = 100$ will have $<1\%$ correction to the power spectrum. We find all models are one to two orders of magnitude below this limit. Therefore, there are no divergent corrections to the power spectrum.}
    \label{fig:Validation:Pspec}
\end{figure}

Figure \ref{fig:Validation:Pspec} shows the power spectrum correction induced by the addition of bispectra to the initial conditions. We compute the correction as 
$\Delta \ln P(k) = (P^{\fNL=+X} + P^{\fNL=-X})/(2P^{\fNL=0}) - 1$, where we drop the function argument, $k$, on the RHS for brevity. This combination suppresses noise, and can be used because the leading order correction to the power spectra scales as $\fNL^2$ and does not cancel when summing contributions from realizations with $\pm \fNL$. Note that this combination cancels cubic-order corrections, $\fNL^3$, but the amplitude of these corrections are significantly lower than those of the quadratic-order terms (which are already small terms). We have checked for the different scalar-exchange models that a naive average using just $+\fNL$ (which does not cancel any noise terms or cubic-order corrections) converges to the noise-suppressed version above, but after using a larger number of realizations.

We find in Figure \ref{appx:Validation:Pk} that the power spectrum correction is minimal in all cases. If a template's fractional change per unit $\fNL$ is below $10^{-4}$, then the overall correction in our simulations (given we use characteristic values of $\fNL = 100$) will be less than $1\%$ to the power spectrum. In our case, all models are at least one order of magnitude below this limit and most models are two orders of magnitude below it. Thus, there is no divergent correction to the power spectrum. This is expected given our multi-pronged approach to building a templates with non-IR divergent terms (Section \ref{sec:sims:ICs} and Appendix \ref{appx:ICs:Divergent}).

\subsection{Trispectrum}\label{appx:Validation:Tk}

Finally, as mentioned in \citetalias{Scoccimarro2012PNGs}, the choice of kernel can impact not only the power spectrum, but also the trispectrum induced into the initial conditions. The method discussed in Section \ref{sec:sims:ICs} guarantees the initial conditions will exhibit a given bispectrum, but it does not explicitly guarantee that the higher-order correlations of the field will be well-behaved. The latter aspect is tested by measuring the trispectrum in these initial conditions and showing they are consistent with noise.\footnote{We follow previous works in assuming that a negligible trispectrum measurement indicates negligible power for all poly-spectra of order higher than the trispectrum \citep[\eg][]{Jung2023fNLHMFQuijote}} We use the estimator discussed in Appendix B2 of \citet{Goldstein:2024:CosmoColl}, which is an extension of the bispectrum estimator of \citet{Scoccimarro:2015:Bispectrum}.

Note that all the bispectrum models we consider (Section \ref{sec:sims:Models}) have nonzero power in the squeezed limit. This naturally generates a trispectrum with amplitude $\tau_{\rm NL} \geq (6f_{\rm NL}/5)^2$ \citep{Suyama:2007bg}. Formally, there is a nonzero trispectrum in our initial conditions simply due to the presence of bispectra.\footnote{Note that this trispectrum contribution scales as $\phi^3 \tau_{\rm NL}$ and therefore has an amplitude that is a factor $\phi \fNL = 10^{-3}$ smaller than the bispectrum contribution, for the characteristic value of $\fNL = 100$.} However, we explicitly verify below that the trispectrum signal in our initial conditions is of negligible amplitude.

\begin{figure}
    \centering
    \includegraphics[width=\columnwidth]{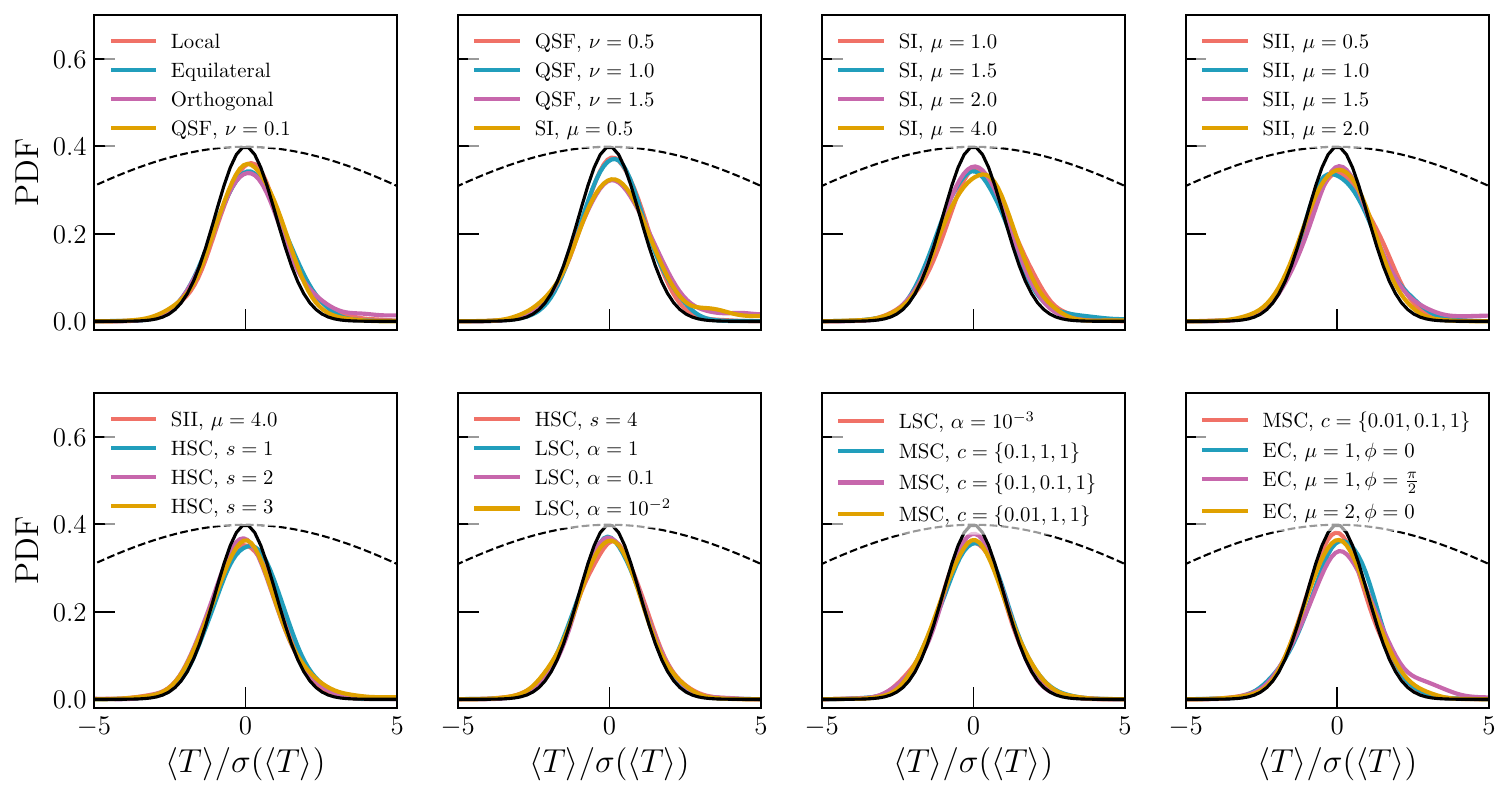}
    \caption{The measurement significance for the average trispectrum (measured over 30 realizations). We measure the significance of the mean trispectrum in each of the 2000 bins and present them as a histogram. A unit Gaussian is overplotted for reference. The measured significances closely match the Gaussian distribution, indicating the mean trispectrum is consistent with noise. We note that we are presenting the significance of the averaged, cosmic variance-suppressed measurement of the trispectrum. The above test is therefore a very conservative search for trispectrum signals in our ICs. The uncertainty for a single realization (not the mean) with no noise suppression is at least a factor of 7 larger (see text for more details). We overplot a Gaussian with width $\sigma = 7$ for reference.}
    \label{fig:Validation:Tk}
\end{figure}

Figure \ref{fig:Validation:Tk} shows the statistical significance of the measured trispectra. In particular, we compute the average trispectrum across 30 realizations of the ICs. We then compute the uncertainty on this mean using $300$ bootstrap realizations. Our estimate is computed as $T^{\fNL = +X} - T^{\fNL = 0}$ and naturally removes the connected trispectrum generated by Gaussian fields. The histograms in Figure \ref{fig:Validation:Tk} are the distribution of the significance values for each of the 2000 elements in the full datavector. We find the distributions are close to Gaussian, suggesting the measurements are indeed consistent with noise. There are some deviations, but these are minor. We stress that the measurement in Figure \ref{fig:Validation:Tk} presents the significance of the \textit{mean, cosmic variance-suppressed, noise-suppressed} trispectrum. We find the actual uncertainty on the measurement in a given trispectrum bin is around seven times larger.\footnote{This is estimated by computing the ratio for each of the 2000 bins and taking the median over them. The median is an intentionally conservative choice for a statistic. When computing the average difference, which accounts for the fact that certain bins have dramatically higher changes, we get ratios of 50 to 100 instead of the value of 7 used in our test.} This distribution is shown in the dashed line and is much wider than the deviations, i.e., the measured mean trispectrum is significantly smaller than the uncertainty of the trispectrum measurement in a single realization of the initial conditions.

We conclude that there is no discernible trispectrum signal in our initial conditions. We follow previous works in generalizing this result to imply our approach does not generate unphysical amplitudes for higher-order poly-spectra beyond the bispectrum \citep{Jung2023fNLHMFQuijote}.

\section{Forward Model of Lensing Field}\label{appx:ForwardModel}

In this work, we are interested in the impact of $\fNL$ as observed by a weak lensing survey. For this, we must construct lensing maps. This can be done by using the density fields of $N$-body simulations to create lensing convergence fields, and then post-processing these fields to match the observed data. Various aspects of these procedures have been utilized in existing analyses/forecasts of weak lensing data \citep{Fluri2019DeepLearningKIDS,  Zurcher2021WLForecast, Fluri2022wCDMKIDS, Gatti2022MomentsDESY3, Zurcher2022WLPeaks, Gatti2023SC, Anbajagane2023CDFs, Anbajagane2023Inflation}. We follow the approach previously detailed in \citet{Anbajagane2023Inflation}, and briefly reproduce the salient points below. The survey we focus on is the Rubin Observatory Legacy Survey of Space and Time (LSST), which is a 14,000 deg$^2$ survey that will have nearly 30 source galaxies per sq. arcmin of sky by the final Year 10 datarelease. We detail below the exact steps in our forward modeling procedure to make lensing fields corresponding to these surveys:

\paragraph{Constructing Lensing Convergence Shells} We start from \textsc{HEALPix} maps of projected density field at different redshifts. These shells can then be converted into the convergence $\kappa$ using Eq.~\ref{eqn:convergence_definition}.

\paragraph{Source Galaxy Redshift Distributions} After obtaining convergence shells at different redshifts, we construct the convergence field for each of the five tomographic bins forecasted for the LSST Y10 dataset. These fields are computed as a weighted average over redshift, where the weights are the source galaxy $n(z)$ of the chosen bin,
\begin{equation}
    \kappa^A(\nhat) = \sum_{j = 1}^{N_{\rm steps}} n^A(z_j)\kappa(\nhat, z_j)\Delta z\,,
\end{equation}
where $\kappa^A$ is the true convergence of a tomographic bin, $A$. The $n(z)$ is the same one utilized in \citet[][see their Table 1]{Zhang2022CMBLSS} and later in \citet[][]{Anbajagane2023Inflation} (see their Figure 1).

\paragraph{Constructing Lensing Shear Shells} Weak lensing surveys observe galaxy shapes that are tracers of the shear field, $\gamma$. The shear can be transformed into the convergence field using the Kaiser-Squires (KS) transform \citep{Kaiser1993KS}, implemented in harmonic space as
\begin{equation}\label{eqn:Kappa2Shear}
    \gamma^{\ell m}_E + i\gamma^{\ell m}_B = -\sqrt{\frac{(\ell + 2)(\ell - 1)}{\ell(\ell + 1)}} \bigg(\kappa^{\ell m}_E + i\kappa^{\ell m}_B \bigg)\,,
\end{equation}
where $X_{\{E, B\}}$ are the E-mode and B-mode (or Q and U polarizations, in \textsc{HealPix} notation) of the field.

\paragraph{Shape Noise} Once the two shear fields are generated, we add the relevant shape noise in real space. This is modelled as a Gaussian with a standard deviation of
\begin{equation}\label{eqn:shapenoise}
    \sigma_\gamma = \frac{\sigma_e}{\sqrt{n_{\rm gal}A_{\rm pix}}}\,,
\end{equation}
where $n_{\rm gal}$ is the source galaxy number density, and $A_{\rm pix}$ is the pixel area for a given map resolution. All maps in this work use $\texttt{NSIDE}=1024$, corresponding to a pixel resolution of $3.2 \arcmin$. The per-galaxy shape noise is taken to be $\sigma_e = 0.26$. 

\paragraph{Survey Mask \& Mass Map Construction} The noisy shear field --- which is the sum of the true shear fields and the shape noise fields --- is then masked according to the survey footprint. We divide the sky into three equal-area cutouts of roughly 14,000 deg$^2$ each, which is the expected area coverage for Y10. The noisy shear maps are converted to convergence using Eq.~\ref{eqn:Kappa2Shear}. We only use the resulting E-mode field, $\kappa_E$, for our analyses. This follows the same procedures used in \citet{Chang2018MassMap, Niall2021MassMap}. The B-mode field is nonzero in this case as the presence of a mask transfers some fraction of power (and thus cosmological information) from E-modes to B-modes. 

\paragraph{Summary} Lensing convergence maps are constructed from the density shells of the simulation. The source galaxy $n(z)$ distribution is used to obtain the convergence map in a given tomographic bin. This convergence map is converted to shear maps, the relevant shape noise is added, the relevant survey mask is applied, and then the noisy, masked shear maps are converted back to a noisy convergence map. The set of procedures listed above is the standard approach for forward modeling the lensing field \citep[\eg][]{Zurcher2021WLForecast, Zurcher2022WLPeaks, Gatti2022MomentsDESY3, Anbajagane2023CDFs, Gatti:2024:WPH, Prat:2025:Homology}. Thus, our final convergence maps will be an accurate representation of the survey data.

A number of known effects have been left out of our forward modeling procedure above: mean redshift uncertainties \citep[\eg][]{Myles2021PhotoZ}, multiplicative bias \citep[\eg][]{MacCrann2022ImsimsY3}, clustering of source galaxies \citep[\eg][]{Krause2021Methods, Gatti2020Moments, Gatti2023SC}, and reduced shear \citep[\eg][]{Krause2010ReducedShear, Gatti2020Moments}, to name a few. This choice has been made for simplicity, and because these factors are not expected to change the Fisher information constraints by a notable amount; either because the effect does not include a nuisance parameter to marginalize over (\eg\, reduced shear) or is an effect with accurate enough calibration such that the nuisance parameter marginalization has a negligible effect on the final constraints (\eg\,  mean redshift, multiplicative bias). The main systematic left out in our analysis is that of intrinsic alignments (IA), which is the correlation of galaxy shapes \citep{Troxel2015IAReview, Lamman2023}. We have ommitted this effect in our work due to numerical considerations (see Section \ref{sec:results:lensing}) and note that marginalizing over IA had negligible effect for the local, equilateral, and orthogonal templates \citep[][see their Figure 2]{Anbajagane2023Inflation}. Moreover, some IA effects show unique signatures in Universes with PNGs so IA can also serve as an additional, novel probe of inflation rather than as a common systematic \citep[\eg][]{Schmidt:2015:IA, Philcox:2024:Shapes}.

\newpage
\section{Additional Bispectrum Templates}\label{appx:templates}

This appendix lists a few additional templates that were omitted in the main text for brevity. The first is the shape resulting from the interaction including spatial derivatives, $(\partial_i \phi)^2 \sigma$, and can be computed as \citep{Sohn:2024:Colliders}

\begin{mybox}
{\small
\begin{align*} \label{eqn:template:SII}
    S_{\rm col.}^{\rm sp} & = \frac{k_3(k_3^2-k_1^2-k_2^2)}{ \beta k_{12}^3} \bigg[ 6-\frac{6\beta k_3}{(\beta+2)k_T} + \frac{2\beta(\beta+1)k_3^2}{ (\beta+2)^2 k_T^2} + \frac{k_1^2+k_2^2}{k_1 k_2}\bigg(2-\frac{\beta k_3}{(\beta+2)k_T}\bigg) \bigg] \bigg(\frac{k_T}{k_{12}}\bigg)^{-\frac{\beta}{\beta+2}} \nonumber\\
    & + \frac{k_3^2-k_1^2-k_2^2}{k_1k_2} \bigg(\frac{k_3}{k_{12}}\bigg)^{\frac{1}{2}} \bigg\{ \sqrt{\frac{\pi^3(\beta+2)}{\mu \sinh(2\pi \mu) }} \cos\bigg[\mu\log\bigg(\frac{k_3}{2k_{12}}\bigg) + \delta_1 \bigg]\nonumber\\
    &+ \frac{k_1k_2}{k_{12}^2}\sqrt{\frac{\pi^3\beta(\beta+2)}{\mu \sinh(2\pi \mu) }} \cos\bigg[\mu\log\bigg(\frac{k_3}{2k_{12}}\bigg) + \delta_2 \bigg] \bigg\}\nonumber\\
    & + \frac{k_3 \left(k_1^2+k_2^2-k_3^2\right)}{12 \cosh(\pi\mu) k_1 k_2 k_{12}^4} \bigg[ 2\left(2 \mu ^4-1\right) k_{12} \bigg((k_1^2+k_2^2+3k_1k_2) \log ^2\left(\frac{k_T}{k_{12}}\right)\nonumber\\
    &+\frac{k_3 }{k_T^2}(k_1^3 + k_2^3 +(k_1^2+k_2^2)k_3 + 7k_1k_2 k_{12}+5k_1k_2k_3) \log \left(\frac{k_T}{k_{12}}\right)+\frac{ k_1 k_2 k_3^2 }{k_T^2} \bigg)\nonumber\\
    & + \mu ^2 k_3 
   \bigg(\frac{k_3}{k_{12}}\bigg)^{\frac{8 \mu ^2}{8 \mu ^2+1}} \left(
   \frac{32 \mu ^2+3}{8 \mu ^2+1} \left(\frac{k_1k_2}{8 \mu ^2+1}  - k_1^2 -k_2^2 -5k_1k_2 \right)\log \left(\frac{k_T}{k_{12}}\right) \right.\nonumber\\[5pt]
   & \left.\left.+\frac{2}{8 \mu ^2+1} \frac{k_1k_2k_3}{k_T}-\frac{k_3}{k_T^2} \Big( k_1^3 + k_2^3 +(k_1^2+k_2^2)k_3 + 11k_1k_2k_{12}+9k_1k_2k_3\Big) \right)\right] + 2~{\rm perm.}\,,
\end{align*}
}
\end{mybox}

\noindent with the two phases $\delta_1 = \arg [ (\frac{3}{2}+i\mu)\Gamma(\frac{1}{2}+i\mu)\Gamma(-i\mu) (i+\text{cosech}\hskip 1pt\pi\mu) ]$ and $\delta_2 = \arg [ \Gamma(\frac{5}{2}+i\mu) \Gamma(-i\mu) (i+\text{cosech}\hskip 1pt\pi\mu) ]$.

Our work also uses the standard primordial templates represented by the local, equilateral, and orthogonal types. For completeness, we quote these here:
\begin{mybox}
\begin{align}
     S^{\rm local} & = \frac{\kThree^2}{\kOne\kTwo}+ {\rm 2~ perm.}\,,\\
     S^{\rm equil} & = \bigg(\frac{\kThree^2}{\kOne\kTwo}+ {\rm 2~ perms.}\bigg) - 2 + \bigg(\frac{\kOne}{\kThree} + {\rm 5~ perms.}\bigg)\,,\\
     S^{\rm ortho} & = -3\bigg(\frac{\kThree^2}{\kOne\kTwo}+ {\rm 2~ perms.}\bigg) - 8 + 3\bigg(\frac{\kOne}{\kThree} + {\rm 5~ perms.}\bigg)\,.
\end{align}
\end{mybox}
As is the case with the other templates, the presented expressions above do not contain the normalization factors needed to ensure the templates follow $S(k, k, k) = 1$. However, we reiterate that the templates are correctly normalized when injected in our simulations. Figure \ref{fig:Template} confirms this by showing the templates are $S(k, k, k) \approx 1$ for all templates. There is a mild scale dependence as the normalization is done for a scale-independent power spectrum (i.e., for $n_s = 1$) so a scale-dependence is induced once we generate templates using the $n_s \neq 1$ power spectrum. 

The above orthogonal template is also known to be an inaccurate approximation to the true behavior of single-field inflation \citep{Senatore2010WMAP5pngs}, as it has an incorrect scaling in the squeezed limit. A more accurate template is provided in Appendix B of \citet{Senatore2010WMAP5pngs} and has been utilized by \citet{Coulton2022QuijotePNG} for generating simulations. In this work, we use the original template for easier comparisons with existing analyses \& constraints \citep[\eg][]{Scoccimarro2012PNGs, Planck:2014:PNGs, Planck:2016:PNGs, Planck2020PNGs}. Both templates are highly correlated with the true shape generated by cubic interactions of the inflaton \citep{Senatore2010WMAP5pngs}.


\label{lastpage}
\end{document}